\documentclass[twoside,11pt,preprint]{article}

\usepackage{blindtext}

\usepackage[abbrvbib, preprint]{dmlr2e}

\usepackage{dmlr2e}
\usepackage{hyperref}
\usepackage{url}
\usepackage{multirow}
\usepackage{booktabs}
\usepackage{graphicx}
\usepackage{tikz}
\usetikzlibrary{positioning, shapes.geometric, arrows.meta, fit}
\usepackage{pgfplots}
\pgfplotsset{compat=1.18}
\usepackage{subcaption}
\usepackage{amsmath}

\usepackage{lastpage}
\dmlrheading{}{Proceedings of the Data-Centric Machine Learning Workshop(DMLR) at ICLR, 2024}{1-\pageref{LastPage}}{}{}{}{} 
\def\openreview{\url{https://openreview.net/forum?id=MmUfbFD8J0&noteId=54ZvqypOrU}} 

\ShortHeadings{Mesh-independent superresolution approach to fluid flow predictions}{Sarkar et al}
\firstpageno{1}

\begin{document}

\title{PointSAGE : Mesh-independent superresolution approach to fluid flow predictions}

\author{\name Rajat Kumar Sarkar \email rajat.sarkar1@tcs.com \\
      \addr Researcher\\
      TCS Research\\
      \AND
      \name Krishna Sai Sudhir Aripirala \email k.aripirala@tcs.com \\
      \addr Researcher\\
      TCS Research\\
      \AND
      \name Vishal Jadhav \email vi.suja@tcs.com \\
      \addr Scientist\\
      TCS Research\\
      \AND
      \name Sagar Srinivas Sakhinana \email sagar.sakhinana@tcs.com \\
      \addr Scientist\\
      TCS Research\\
      \AND
      \name Venkataramana Runkana \email venkat.runkana@tcs.com \\
      \addr Chief Scientist\\
      TCS Research\\
      }

\maketitle

\begin{abstract}
Computational Fluid Dynamics (CFD) serves as a powerful tool for simulating fluid flow across diverse industries. High-resolution CFD simulations offer valuable insights into fluid behavior and flow patterns, aiding in optimizing design features or enhancing system performance. However, as resolution increases, computational data requirements and time increase proportionately. This presents a persistent challenge in CFD. Recently, efforts have been directed towards accurately predicting fine-mesh simulations using coarse-mesh simulations, with geometry and boundary conditions as input. Drawing inspiration from models designed for super-resolution, deep learning techniques like UNets have been applied to address this challenge. However, these existing methods are limited to structured data and fail if the mesh is unstructured due to its inability to convolute. Additionally, incorporating geometry/mesh information in the training process introduces drawbacks such as increased data requirements, challenges in generalizing to unseen geometries for the same physical phenomena, and issues with robustness to mesh distortions. To address these concerns, we propose a novel framework, \textbf{PointSAGE} a mesh-independent network that leverages the unordered, mesh-less nature of Pointcloud to learn the complex fluid flow and directly predict fine simulations, completely neglecting mesh information. Utilizing an adaptable framework, the model accurately predicts the fine data across diverse point cloud sizes, \textit{regardless of the training dataset's dimension}. We have evaluated the effectiveness of PointSAGE on diverse datasets in different scenarios, demonstrating notable results and a significant acceleration in computational time in generating fine simulations compared to standard CFD techniques.  

\end{abstract}


\section{Introduction}
The Navier-Stokes equation stands as a fundamental cornerstone, providing insight into the complex physics that governs scientific and engineering phenomena. However, its intrinsic non-linearity poses challenges. In response, Computational Fluid Dynamics (CFD) has emerged, employing various computational methods to tackle fluid flow complexities. Efforts to predict intricate flows with precision often demand fine resolutions, intensifying computational requirements. In the typical scenario of Direct Numerical Simulation (DNS), fine simulations require a large number of CPU hours and approximately a terabyte (TB) of memory for data storage (\cite{hawkes2005direct}). In industrial and real-world contexts, flow dynamics often display significant computational complexity, characterized by turbulence (\textit{pipeline flow}), multi-physics (\textit{aerospace applications}), and multi-phase behaviors (\textit{combustion reactors}). Consequently, simulating these phenomena require extensive computational resources and data storage demands. In contrast, coarse simulations would consume only half the time needed for fine simulations, with memory requirements reduced to about 1/100th for simulating the same physical phenomena.

Hence, coarse grid simulations have become important due to computational efficiency. Yet, the persistent pursuit of understanding complex phenomena drives the ongoing demand for fine-mesh simulations. Recent advancements inspired by super-resolution techniques have introduced deep-learning methodologies for predicting fine-mesh simulations from coarse-mesh counterparts. Utilizing established architectures like MLP (\cite{erichson2020shallow, nair2020leveraging}, U-Nets(\cite{sarkar2023redefining, pathak2020using}), and GANs (\cite{xie2018tempogan, bode2019deep, kim2021unsupervised, guemes2021coarse, yousif2021high, bode2021using}), GNNs(\cite{pfaff2020learning}) these approaches show promise in overcoming computational challenges to generate fine simulations.

Despite these advancements, current research faces certain challenges. These models adhere to the conventional definition of super-resolution, wherein the fine simulation is \textit{coarsened} with the aid of down-sampling techniques such as max-pooling and nearest neighbor methods (\cite{gao2021super, bode2019deep, bode2021using, esmaeilzadeh2020meshfreeflownet}). Consequently, the model is trained to learn the mapping between the fine data and its down-sampled counterpart, enabling it to reconstruct the fine data from the down-sampled version. Since this approach retains the physics in the down-sampling process, the training inherently carries a slight bias, resulting in accurate outcomes (\cite{sarkar2023redefining}). However, real-life scenarios often involve coarse data that is not directly down-sampled from the fine data. As a consequence, these models struggle to predict accurately across different contexts. Recently, researchers have begun utilizing actual coarse data, rather than relying solely on down-sampled fine data, which aligns more closely with real-life scenarios.

However, the current research primarily focuses on structured and regular data, limiting the model's adaptability to various data formats, including unstructured or irregular data (\cite{sarkar2023redefining, pathak2020using}). When confronted with unstructured data, these models fail due to their inability to undergo convolution. Additionally, integrating mesh information during the training phase poses numerous challenges, such as increased data requirements and complexity, leading to extended training time. Training the model on specific geometries restricts its predictive capabilities, especially when assessing unseen or novel geometries for the same physical phenomena, resulting in generalization issues. Moreover, obtaining accurate mesh information in real-world scenarios presents a practical challenge. Consequently, the model's robustness is compromised, affecting its ability to precisely predict fine-mesh simulations in real-world applications.

Consequently, there is a pressing need for a framework that isn't reliant on the geometry or mesh information of the domain. Point clouds offer a comprehensive representation of 3D space by capturing the spatial coordinates of individual points within the domain. In recent times, they have gained prominence due to their ability to handle unstructured data, thanks to their unordered nature (\cite{qi2017pointnet}). This flexibility allows point clouds to capture intricate details and irregularities, making them well-suited for modeling diverse and complex scenarios.

To address these challenges, we introduced \textbf{PointSAGE}, a novel \textit{mesh-independent} framework that disregards mesh information by representing data as point clouds sourced from various entities such as cell centers and nodes in the computational domain. This approach affords us the flexibility to manage irregular and unstructured data owing to its unordered nature. PointSAGE incorporates the global feature extraction approach inspired by the "classification network" in the \textit{PointNet} architecture (\cite{qi2017pointnet}), and leverages the potency of \textit{SAGEConv} (\cite{hamilton2018inductive}) to capture the local inter-dependencies of features within the fluid flow. The model is developed to seamlessly predict fine mesh data, independent of the size/dimensionality of the training sets, \textit{thereby enabling accurate prediction for any value of \(n\)}.  Through comprehensive testing across diverse datasets and scenarios, PointSAGE demonstrated remarkable performance, showcasing significant reductions in training time compared to state-of-the-art (SOTA) techniques. Additionally, our model exhibited enhanced computational efficiency, both in terms of \textit{memory usage and processing time}, when contrasted with conventional CFD methods for fine simulation generation.

\section{PointSAGE Super-resolution on Point Cloud}
\begin{figure}[htbp]
    \centering
    \includegraphics[width=\linewidth, trim={1cm 1.9cm 1cm 2.15cm}, clip]{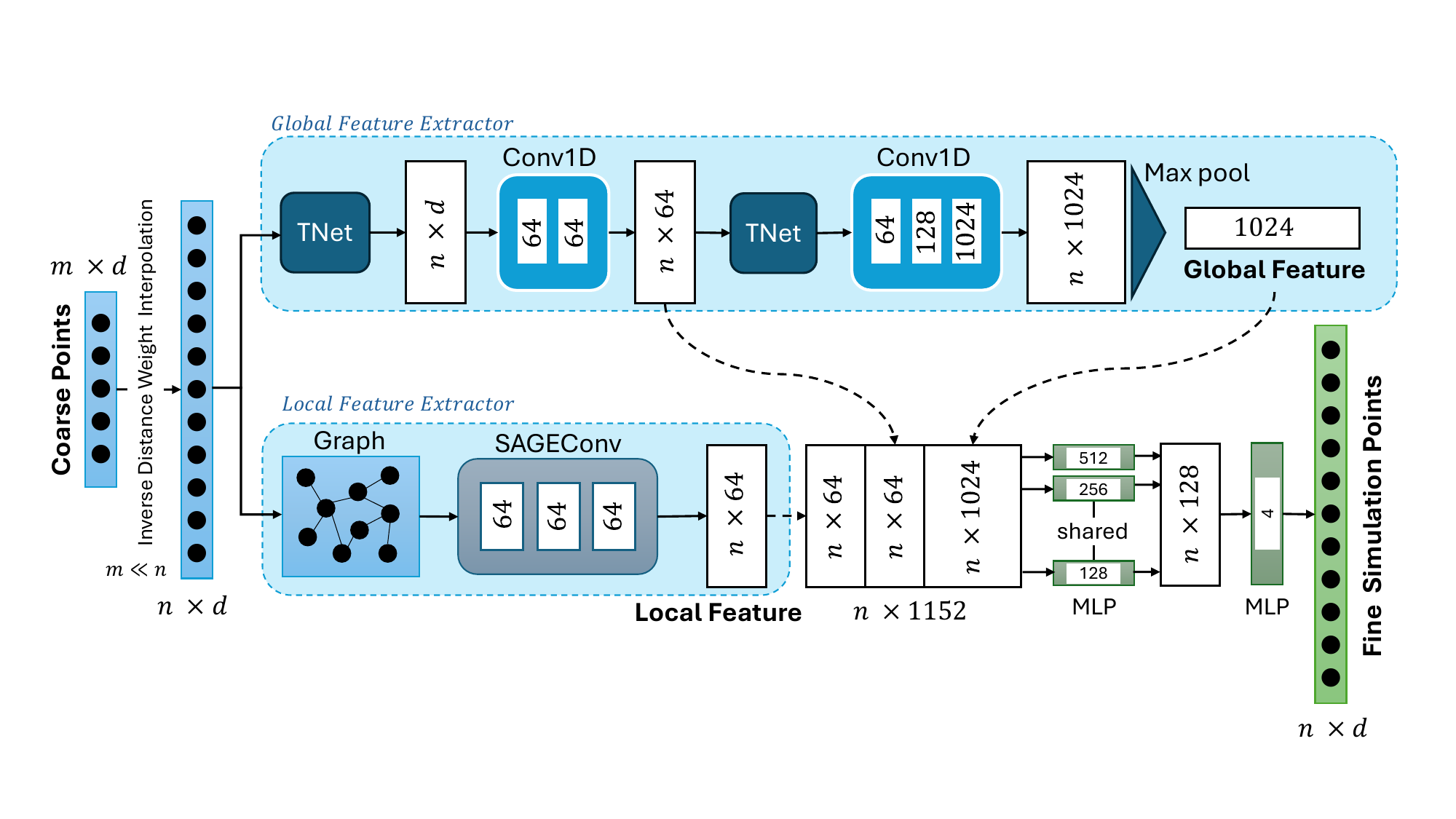}
    \caption{\textbf{PointSAGE Architecture} : The coarse data undergoes dimension matching with the fine data through the IDW Up-sampler technique. Subsequently, the up-sampled coarse data is concurrently processed by two modules: the Global (\textit{PointNet}) and Local (\textit{GraphSAGE}) feature extractors. These outputs are then fused through concatenation to accurately predict the fine mesh data.}
    \label{fig:PointSAGE}
\end{figure}

In this section, we present the proposed architecture for super-resolving the point cloud of coarse simulation data to match the fine simulation data. The training dataset comprises pairs of coarse mesh data and fine mesh data represented as \textbf{point clouds}, denoted as \((C, F)\). The model \(f: C \rightarrow F\) is crafted to accurately capture and map the non-linear relationship between the coarse point cloud \(C \in \mathbb{R}^{m \times d}\) and the fine point cloud \(F \in \mathbb{R}^{n \times d}\), where \(m\) and \(n\) denote the number of points, and \(d\) represents the number of features, with \(m \ll n\). A crucial aspect of our approach is our independence from mesh information, solely concentrating on gathering state-variable information at each point cloud. As illustrated in Figure \ref{fig:PointSAGE}, our architecture comprises three components: (a) an Inverse Distance Weighting (IDW) upsampler, (b) a Global Feature Extractor, and (c) a Local Feature Extractor.

\subsection{IDW Upsampler}
The coarse point cloud \( m \) undergoes an up-sampling process to enhance its resolution, aligning it with the shape of the fine-point cloud \( n \). For upsampling, we employ Inverse Distance Weighting with exponential weights. In this interpolated feature \( V(n) \) of the coarse mesh at point \( n \), we calculate it as:
\[
V(n) = \frac{\sum_{i=1}^{n} V_i \cdot e^{-(\frac{D_i}{D_0})^2}}{\sum_{i=1}^{n} e^{-(\frac{D_i}{D_0})^2}}
\]
where \( D_i \) represents the Euclidean distance between coarse points \(m\) and fine points \(n\), and \(D_0\) is the correlation distance (in this case, the maximum distance is used). After the upsampling, we feed the interpolated point cloud into two other architecture components, as shown in Figure \ref{fig:PointSAGE}.
\subsection{Global Feature Extractor}
The interpolated point cloud \(V \in \mathbb{R}^{n \times d}\) undergoes an input transformation using a TNet of dimension \(d\). This transformation, achieved through matrix multiplication with TNet output, enables the network to learn robust feature representations invariant to geometric transformations. Subsequently, two Conv1D layers with hidden channels of 64 are employed to extract learned features from the input feature dimension to the hidden dimension. Following this, feature transformation is performed using another TNet of dimension 64 with matrix multiplication, facilitating the capture of complex patterns and structures present in the data. Multiple Conv1D layers with a max-pooling layer are utilized to aggregate features from all points, extracting a global feature vector of dimension 1024 that represents the entire point cloud. This entire process draws inspiration from the classification section of PointNet.

\subsection{Local Feature Extractor}
Simultaneously, the interpolated point cloud undergoes local feature extractor along with the global feature extractor. In this component, we construct a graph \(G = (P,E)\), where \(P\) represents the vertices of the graph generated from the \(n\) interpolated point cloud, and \(E\) denotes the edges connecting the vertices to their neighbors within a fixed radius \(r\). The graph construction is achieved using a radius near-neighbor technique with a radius \(r = 0.005\), and a maximum of 32 neighbors is given by
\[
E = \{(p_i, p_j) \mid \| \mathbf{x}_i - \mathbf{x}_j \|_2 < r\}
\]
Subsequently, the constructed graph undergoes multiple SAGEConv layers with a hidden dimension of 64, which aggregate information from neighboring nodes in the graph. This allows for capturing local geometric structures and relationships, facilitating effective local feature learning and extraction of the local feature vector.

The local feature vector is then concatenated with the global feature vector and passed through multiple multi-layer perceptrons (MLPs) to effectively map the coarse point cloud \(C\) to the fine point cloud \(F\). The Mean Squared Error (MSE) is employed as the loss function for training the model. Importantly, our methodology deliberately avoids incorporating any mesh information throughout this entire process. This ensures that our model remains entirely mesh-independent, contributing to its robustness and versatility in handling diverse mesh representations such as irregularly structured and unstructured grids.

\section{Results and Discussion}
To demonstrate the effectiveness of our model, we conducted experiments on three diverse datasets: a forward-facing step, a lid-driven cavity, and methane combustion simulations. We evaluated our model's performance using 2D point cloud data from the forward-facing step simulation and extended it to 3D lid-driven cavity simulation data. Additionally, we compared our model to recent work, specifically PIUNet, \cite{sarkar2023redefining}, which operates on regular grid fluid flow prediction, using 2D methane combustion data. This comparison aims to highlight the advantages of our approach, PointSAGE, utilizing point cloud data from the simulation.

\textbf{Forward-Step:} The transient CFD simulation of the forward step provides valuable insights into the complex flow phenomena associated with separated and reattached flows in a step-like configuration. We specifically focused on predicting the shock wave generated by supersonic flow at the inlet to a rectangular geometry with a step near the inlet region. Details of the CFD simulation are explained in the appendix \ref{sec:Case_study_1}. To showcase our approach's effectiveness, we demonstrated the model's performance in two scenarios. In the first scenario, we varied the inlet velocity (\(U_{\infty} \in [2,5]\) m/s) for a given Aspect Ratio (AR) of 3. In the second scenario, we varied both the inlet velocity (\(U_{\infty} \in [2,5]\) m/s) and the Aspect Ratio (\(AR \in [3,6]\)). Table \ref{tab:forward} shows that PointSAGE outperforms other mesh-independent approaches in both scenarios. The problem involves four features: \(x\)-direction velocity (\(U_x\)), \(y\)-direction velocity (\(U_y\)), pressure (\(Pa\)), and Mach Number (\(Ma\)). In terms of RMSE, we achieved approximately 70$\%$ and 20$\%$ enhancement in all features compared to the second-best results in scenarios 1 and 2, respectively.

\begin{table}[htbp]
\centering
\caption{Performance of PointSAGE fine mesh prediction on 2D forward step dataset}
\label{tab:forward}
\setlength{\tabcolsep}{0.7\tabcolsep} 
\resizebox{\textwidth}{!}{%
\begin{tabular}{@{}ccccccccccccccccc@{}}
\toprule
\multicolumn{17}{c}{\textbf{Scenario 1:} Varying inlet velocity for an Aspect Ratio (AR) of 3 (with a 2.4m block length beyond the step)(3352 pts $\rightarrow$ 42702 pts)} \\ \midrule
\multicolumn{1}{c|}{\multirow{4}{*}{\rotatebox[origin=c]{90}{U$_x$}}} &
  \multicolumn{1}{c|}{\textbf{Algorithm}} &
  \textbf{MAE} &
  \textbf{RMSE} &
  \multicolumn{1}{c|}{\textbf{R$^{2}$}} &
  \multicolumn{1}{c|}{\multirow{4}{*}{\rotatebox[origin=c]{90}{U$_y$}}} &
  \textbf{MAE} &
  \textbf{RMSE} &
  \multicolumn{1}{c|}{\textbf{R$^{2}$}} &
  \multicolumn{1}{c|}{\multirow{4}{*}{\rotatebox[origin=c]{90}{Pressure}}} &
  \textbf{MAE} &
  \textbf{RMSE} &
  \multicolumn{1}{c|}{\textbf{R$^{2}$}} &
  \multicolumn{1}{c|}{\multirow{4}{*}{\rotatebox[origin=c]{90}{Mach No.}}} &
  \textbf{MAE} &
  \textbf{RMSE} &
  \textbf{R$^{2}$} \\ \cmidrule(lr){2-5} \cmidrule(lr){7-9} \cmidrule(lr){11-13} \cmidrule(l){15-17} 
\multicolumn{1}{c|}{} &
  \multicolumn{1}{c|}{SAGEConv} &
  0.9025 &
  1.0919 &
  \multicolumn{1}{c|}{0.1695} &
  \multicolumn{1}{c|}{} &
  0.2434 &
  0.3847 &
  \multicolumn{1}{c|}{0.5572} &
  \multicolumn{1}{c|}{} &
  2.4264 &
  4.5794 &
  \multicolumn{1}{c|}{0.1626} &
  \multicolumn{1}{c|}{} &
  0.5665 &
  0.7725 &
  0.6726 \\
\multicolumn{1}{c|}{} &
  \multicolumn{1}{c|}{PointNET} &
  0.6428 &
  1.017 &
  \multicolumn{1}{c|}{0.3126} &
  \multicolumn{1}{c|}{} &
  0.2687 &
  0.4065 &
  \multicolumn{1}{c|}{0.4973} &
  \multicolumn{1}{c|}{} &
  2.0413 &
  4.0606 &
  \multicolumn{1}{c|}{0.3416} &
  \multicolumn{1}{c|}{} &
  0.5101 &
  0.758 &
  0.6784 \\
\multicolumn{1}{c|}{} &
  \multicolumn{1}{c|}{\textbf{PointSAGE}} &
  \textbf{0.1651} &
  \textbf{0.3418} &
  \multicolumn{1}{c|}{\textbf{0.8733}} &
  \multicolumn{1}{c|}{} &
  \textbf{0.0941} &
  \textbf{0.1668} &
  \multicolumn{1}{c|}{\textbf{0.8996}} &
  \multicolumn{1}{c|}{} &
  \textbf{0.7323} &
  \textbf{1.6079} &
  \multicolumn{1}{c|}{\textbf{0.8691}} &
  \multicolumn{1}{c|}{} &
  \textbf{0.1476} &
  \textbf{0.275} &
  \textbf{0.9415} \\ \midrule
\multicolumn{17}{c}{\textbf{Scenario 2:} Training on AR 3 and 4, validating on AR 5, testing on AR 6 (6082 pts $\rightarrow$ 91302 pts), with varying inlet velocity and aspect ratio.} \\ \midrule
\multicolumn{1}{c|}{\multirow{4}{*}{\rotatebox[origin=c]{90}{U$_x$}}} &
  \multicolumn{1}{c|}{\textbf{Algorithm}} &
  \textbf{MAE} &
  \textbf{RMSE} &
  \multicolumn{1}{c|}{\textbf{R$^{2}$}} &
  \multicolumn{1}{c|}{\multirow{4}{*}{\rotatebox[origin=c]{90}{U$_y$}}} &
  \textbf{MAE} &
  \textbf{RMSE} &
  \multicolumn{1}{c|}{\textbf{R$^{2}$}} &
  \multicolumn{1}{c|}{\multirow{4}{*}{\rotatebox[origin=c]{90}{Pressure}}}&
  \textbf{MAE} &
  \textbf{RMSE} &
  \multicolumn{1}{c|}{\textbf{R$^{2}$}} &
  \multicolumn{1}{c|}{\multirow{4}{*}{\rotatebox[origin=c]{90}{Mach No.}}} &
  \textbf{MAE} &
  \textbf{RMSE} &
  \textbf{R$^{2}$} \\ \cmidrule(lr){2-5} \cmidrule(lr){7-9} \cmidrule(lr){11-13} \cmidrule(l){15-17} 
\multicolumn{1}{c|}{} &
  \multicolumn{1}{c|}{SAGEConv} &
  0.6254 &
  0.8411 &
  \multicolumn{1}{c|}{-0.9743} &
  \multicolumn{1}{c|}{} &
  0.2324 &
  0.3138 &
  \multicolumn{1}{c|}{-0.4679} &
  \multicolumn{1}{c|}{} &
  1.6785 &
  2.6554 &
  \multicolumn{1}{c|}{-0.3859} &
  \multicolumn{1}{c|}{} &
  0.667 &
  0.8497 &
  -0.8944 \\
\multicolumn{1}{c|}{} &
  \multicolumn{1}{c|}{PointNET} &
  0.5009 &
  0.7747 &
  \multicolumn{1}{c|}{-0.2627} &
  \multicolumn{1}{c|}{} &
  0.2052 &
  0.2985 &
  \multicolumn{1}{c|}{-0.1746} &
  \multicolumn{1}{c|}{} &
  1.3818 &
  2.2660 &
  \multicolumn{1}{c|}{0.0150} &
  \multicolumn{1}{c|}{} &
  0.5147 &
  0.7196 &
  -0.0550 \\
\multicolumn{1}{c|}{} &
  \multicolumn{1}{c|}{\textbf{PointSAGE}} &
  \textbf{0.2810} &
  \textbf{0.5008} &
  \multicolumn{1}{c|}{\textbf{0.3125}} &
  \multicolumn{1}{c|}{} &
  \textbf{0.2007} &
  \textbf{0.2859} &
  \multicolumn{1}{c|}{\textbf{-0.0962}} &
  \multicolumn{1}{c|}{} &
  \textbf{1.1817} &
  \textbf{2.0587} &
  \multicolumn{1}{c|}{\textbf{0.0380}} &
  \multicolumn{1}{c|}{} &
  \textbf{0.2854} &
  \textbf{0.4610} &
  \textbf{0.4408} \\ \bottomrule
\end{tabular}%
}
\end{table}

\begin{figure}[htbp]
    \centering
    \includegraphics[width=1.03\linewidth, trim={0.2cm 0.2cm 0cm 0cm}, clip]{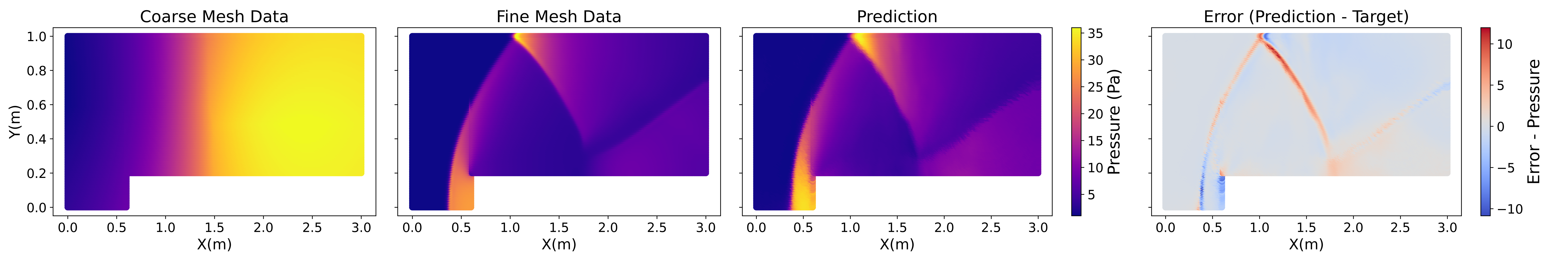}
    \includegraphics[width=1.03\linewidth, trim={0.2cm 0.2cm 0cm 0cm}, clip]{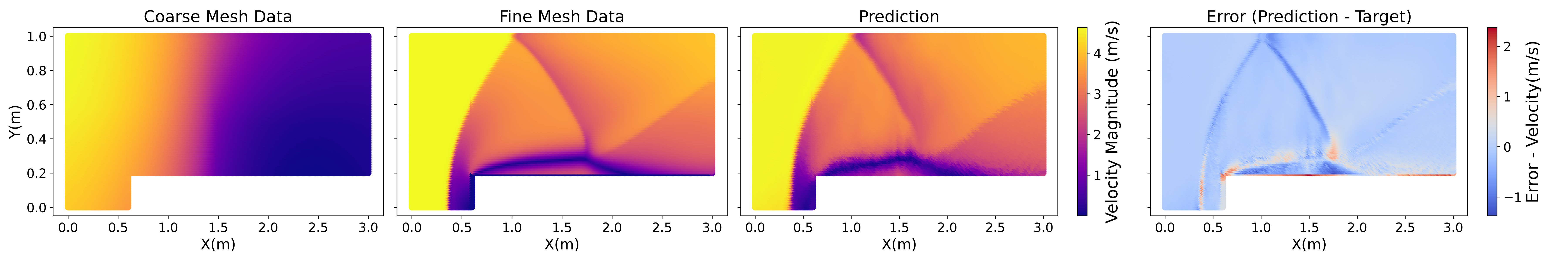}
    \caption{Scenario 1: Pressure and Velocity prediction at t= 3.5s for inlet velocity 4.465 m/s for an AR 3 dataset.}
    \label{fig:Scenario_1}
\end{figure}

From Figure \ref{fig:Scenario_1}, it is evident that at time step \(t = 3.5\) s, there is a noticeable generation of a shock wave near the forward step due to an inlet velocity of \(4.465\) m/s (equivalent to Mach \(4\) in this fluid medium where the velocity of sound is \(a = \sqrt{\gamma R T}\) = \(1\) m/s). Subsequently, after traveling a certain distance, the shock wave gets reflected from the upper surface, creating a reflected shock. Our model, PointSAGE, successfully captures all these phenomena with a maximum absolute pressure error \(E\) of \(10\) in the reflected shock region, where \(E = |\hat{X} - X|\), and \(\hat{X}\) refers to the prediction while \(X\) refers to the fine mesh data.

\begin{figure}
    \centering
    \includegraphics[width=0.9\linewidth, trim={0.2cm 1cm 0cm 2cm}, clip]{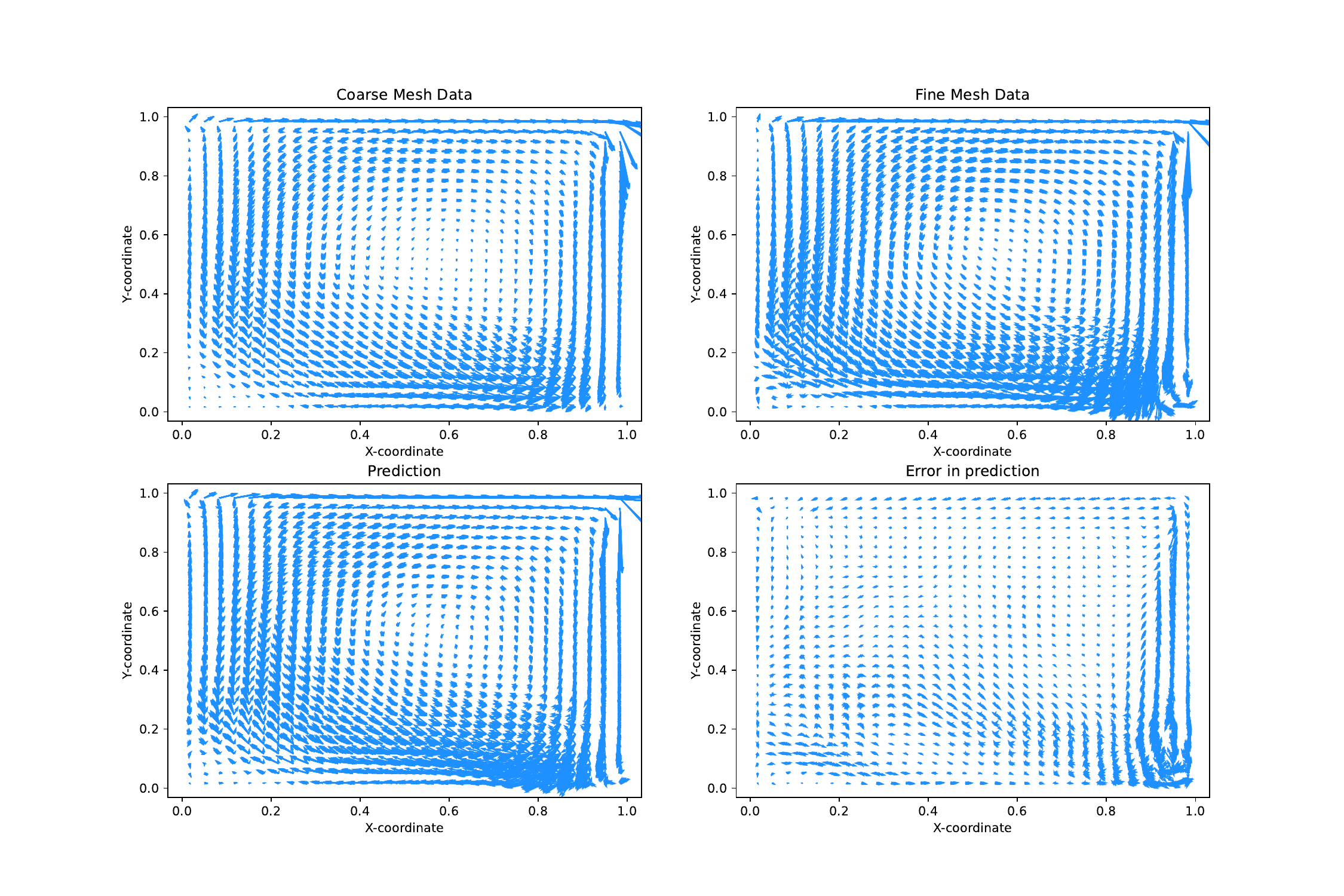}
    \caption{Scenario 1: Velocity prediction for the case of Reynolds number interpolation}
    \label{fig:LDC_Scenario_1}
\end{figure}

\textbf{Lid-Driven Cavity:} \cite{hanna2017coarsegrid} This case study on the lid-driven cavity aims to showcase our model's predictive capabilities. It demonstrates two key aspects: Firstly, the model's ability to learn and predict the turbulence aspect of the flow, notably the bottom-right vortex in the cavity, as depicted in Figure \ref{fig:LDC_Scenario_1}, where the velocity contour is plotted. Secondly, its effectiveness in handling unseen geometries or conditions after training on various scenarios, as evidenced by Table \ref{LDC}. Our model achieves comparable accuracy with existing benchmark techniques in scenarios such as Re interpolation or Re\&GS extrapolation, all while requiring significantly less training time— notably, five times faster. In addition to our developed model, we have also utilized other methods such as SAGEConv and PointNet to demonstrate the effectiveness of these mesh-independent approaches in learning and predicting fine-mesh simulations with notable accuracy. While these methods excel in simpler datasets like the one presented here, our model excels in managing more intricate data scenarios.

\begin{table}
\centering
\renewcommand{\arraystretch}{1.3} 
\setlength{\tabcolsep}{0.6\tabcolsep} 
\caption{Performance of PointSAGE fine mesh prediction on 3D Lid Driven cavity dataset (where Time is Training time)}
\label{tab:lid_driven_cavity}
\resizebox{\textwidth}{!}{%
\begin{tabular}{@{}c|c|c|c|c|c|c|c|c|c|c|c|c|c|c|c|c|c@{}}
\toprule
\multirow{6}{*}{\textbf{\rotatebox[origin=c]{90}{Re Interpolation}}} &
\textbf{\rotatebox[origin=c]{90}{Features}} &
\textbf{Algorithms} &
\textbf{\begin{tabular}[c]{@{}c@{}}MSE \\ (1$e^{-4}$)\end{tabular}} &
\textbf{R$^{2}$} &
\textbf{\begin{tabular}[c]{@{}c@{}}Time \\  (sec)\end{tabular}} &
\multirow{6}{*}{\textbf{\rotatebox[origin=c]{90}{Re Extrapolation}}} &
\textbf{\begin{tabular}[c]{@{}c@{}}MSE \\ (1$e^{-4}$)\end{tabular}} &
\textbf{R$^{2}$} &
\textbf{\begin{tabular}[c]{@{}c@{}}Time \\(sec)\end{tabular}} &
\multirow{6}{*}{\textbf{\rotatebox[origin=c]{90}{  Re $\&$ GS Interpolation}}} &
\textbf{\begin{tabular}[c]{@{}c@{}}MSE\\  (1$e^{-4}$)\end{tabular}} &
\textbf{R$^{2}$} &
\textbf{\begin{tabular}[c]{@{}c@{}}Time\\  (sec)\end{tabular}} &
\multirow{6}{*}{\rotatebox[origin=c]{90}{\textbf{  Re $\&$ GS Extrapolation}}} &
\textbf{\begin{tabular}[c]{@{}c@{}}MSE\\  (1$e^{-4}$)\end{tabular}} &
\textbf{R$^{2}$} &
\textbf{\begin{tabular}[c]{@{}c@{}}Time \\ (sec)\end{tabular}} \\ \cmidrule(lr){2-6} \cmidrule(lr){8-10} \cmidrule(lr){12-14} \cmidrule(l){16-18} 
 & \rotatebox[origin=c]{90}{$U_x$}       & CG-CFD                & 1    & 0.915 & 660 &  & 1   & -     & 660 &  & 1   & -     & 660 &  & 1   & -     & 780 \\ \cmidrule(lr){2-6} \cmidrule(lr){8-10} \cmidrule(lr){12-14} \cmidrule(l){16-18}  
 & \multirow{4}{*}{\rotatebox[origin=c]{90}{$U_x$,$U_y$,$U_z$}}             & UNet                  & 1.3  & 0.971     & 600 &  & 1.4 & 0.944    & 660 &  & -   & -     & -   &  & -   & -     & - \\  
 &                           & SAGEConv             & 2.67 & 0.965 & 80  &  & 3.6 & 0.933 & 83  &  & 3   & 0.959 & 93  &  & 1.7 & 0.978 & 41  \\
 &                           & PointNET              & 2.7  & 0.967 & 26  &  & 3.3 & 0.937 & 29  &  & 3.1 & 0.961 & 28  &  & 1.7 & 0.98  & 13  \\  
 &                           & \textbf{PointSAGE} & 2.6  & 0.959 & 120 &  & 3.5 & 0.924 & 120 &  & 3   & 0.952 & 156 &  & 2.3 & 0.968 & 50  \\ \bottomrule
\end{tabular}%
}
\label{LDC}
\end{table}
\textbf{Methane Combustion:} \cite{yang2019reactingfoam} This case study on methane combustion aims to highlight the contrast between two methodologies: PointSAGE and recent work as mentioned above, PIUNet, which relies solely on a regular mesh grid. We assess various outputs, including the adiabatic flame temperature (\(T_{adia}\)), $x$-direction velocity (\(U_x\)), $y$-direction velocity (\(U_y\)), and mass fractions of species \(CH_4\), \(O_2\), and \(CO_2\). Analyzing the data from Table \ref{tab:methane_combustion}, we observe that PointSAGE achieves comparable results to PIUNet in terms of adiabatic temperature. However, our model outperforms or ranks second best in other features compared to benchmark algorithms. These findings demonstrate that not only does PointSAGE offer a mesh-independent approach, facilitating super-resolution on any mesh or geometry, but it also predicts finer mesh simulation results with comparable accuracy.

\begin{table}[htbp]
\centering
\caption{Performance of PointSAGE fine mesh prediction (1000 pts $\rightarrow$ 50000 pts) on 2D Methane Combustion dataset}
\label{tab:methane_combustion}
\resizebox{\textwidth}{!}{%
\begin{tabular}{@{}c|c|c|ccc|c|ccc|c|ccc@{}}
\toprule
 &
   &
  \textbf{Algorithm} &
  \textbf{MAE} &
  \textbf{RMSE} &
  \textbf{R$^{2}$} &
  \textbf{} &
  \textbf{MAE} &
  \textbf{RMSE} &
  \textbf{R$^{2}$} &
  \textbf{} &
  \textbf{MAE} &
  \textbf{RMSE} &
  \textbf{R$^{2}$} \\ \midrule
\multirow{5}{*}{\textbf{\rotatebox[origin=c]{90}{\begin{tabular}[c]{@{}c@{}}Fluid \\ properties\end{tabular}}}} &
  \multirow{5}{*}{\rotatebox[origin=c]{90}{Temperature}} &
  UNet &
  13.224 &
  30.718 &
  0.9963 &
  \multirow{5}{*}{\rotatebox[origin=c]{90}{$U_x$}} &
  0.0177 &
  0.0296 &
  0.9839 &
  \multirow{5}{*}{\rotatebox[origin=c]{90}{$U_y$}} &
  0.0152 &
  0.0324 &
  0.9830 \\
 &  & PIUNet    & \textbf{10.385} & \textbf{20.954} & \textbf{0.9984} &  & 0.0164 & 0.0286 & \textbf{0.9862} &  & 0.0158 & 0.0324 & 0.9835          \\
 &  & SAGEConv & 18.209          & 30.075          & 0.9959 &  & 0.0079 & 0.0107 & 0.9924          &  & 0.0109 & 0.0153 & \textbf{0.9940}          \\
 &  & PointNET  & 27.543          & 46.745          & 0.9847 &  & 0.0075 & 0.0105 & 0.9836          &  & 0.0154 & 0.0222 & 0.9735          \\
 &
   &
  \textbf{PointSAGE} &
  17.296 &
  28.701 &
  0.9947 &
   &
  \textbf{0.0071} &
  \textbf{0.0105} &
  0.9855 &
   &
  \textbf{0.0104} &
  \textbf{0.0151} &
  0.9930 \\ \midrule
\multirow{5}{*}{\textbf{\rotatebox[origin=c]{90}{\begin{tabular}[c]{@{}c@{}}Mass \\ Fraction\end{tabular}}}} &
  \multirow{5}{*}{\rotatebox[origin=c]{90}{$CH_4$}} &
  UNet &
  0.0140 &
  0.0140 &
  0.9938 &
  \multirow{5}{*}{\rotatebox[origin=c]{90}{$O_2$}} &
  0.0059 &
  0.0116 &
  0.9870 &
  \multirow{5}{*}{\rotatebox[origin=c]{90}{$CO_2$}} &
  0.0058 &
  0.0094 &
  0.9564 \\
 &  & PIUNet    & 0.0138          & 0.0138          & 0.9954 &  & 0.0030 & 0.0106 & 0.9888          &  & 0.0018 & 0.0051 & \textbf{0.9844} \\
 &  & SAGEConv & \textbf{0.0089}          & \textbf{0.0128}          & \textbf{0.9987} &  & 0.0024 & 0.0041 & 0.9974          &  & 0.0013 & 0.0024 & 0.9639          \\
 &  & PointNET  & 0.0125          & 0.0191          & 0.9932 &  & 0.0028 & 0.0053 & 0.9954          &  & 0.0016 & 0.0033 & -386.93         \\
 &
   &
  \textbf{PointSAGE} &
  0.0100 &
  0.0140 &
  0.9971 &
   &
  \textbf{0.0024} &
  \textbf{0.0038} &
  \textbf{0.9975} &
   &
  \textbf{0.0012} &
  \textbf{0.0022} &
  0.9344 \\ \bottomrule
\end{tabular}%
}
\end{table}

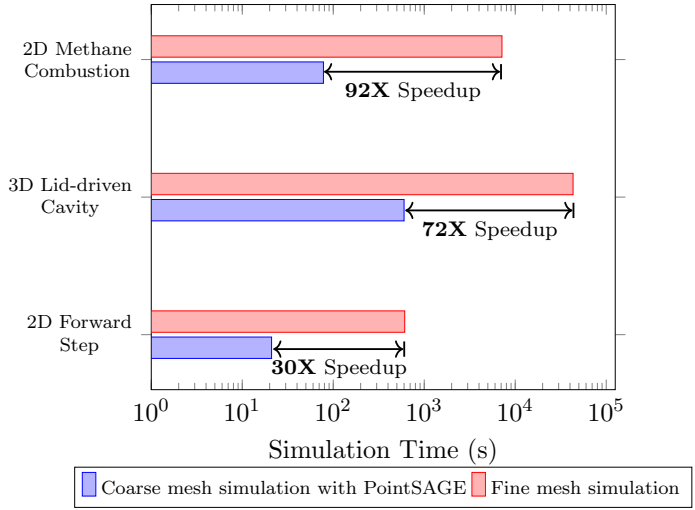
\begin{figure}[htbp]
    \centering
    \begin{tikzpicture}[scale=0.9] 
        \begin{axis}[
            xmin=1, 
            xlabel={Simulation Time (s)}, 
            xmode=log, 
            ytick=data,
            yticklabels={2D Forward \\Step, 3D Lid-driven \\Cavity, 2D Methane \\Combustion},
            yticklabel style={align=center, font=\scriptsize}, 
            legend style={at={(0.5,-0.2)}, anchor=north, legend columns=-1, font=\scriptsize}, 
            enlarge y limits={0.2}, 
            xbar,
            bar width=9pt,
            legend image code/.code={
                \draw [#1] (0cm,-0.1cm) rectangle (0.2cm,0.25cm); },
            ]
            \addplot coordinates {(21,0) (600,1) (78,2)};
            \addplot coordinates {(610,0) (43200,1) (7146.67,2)};
            \legend{Coarse mesh simulation with PointSAGE, Fine mesh simulation}
        \end{axis}
        \draw[<->|, line width=0.7pt] (1.8,0.6) -- (3.75,0.6) node[midway,below, font=\scriptsize] {\textbf{30X} Speedup};
        \draw[<->|, line width=0.7pt] (3.75,2.65) -- (6.25,2.65) node[midway,below, font=\scriptsize] {\textbf{72X} Speedup};
        \draw[<->|, line width=0.7pt] (2.55,4.7) -- (5.18,4.7) node[midway,below, font=\scriptsize] {\textbf{92X} Speedup};
    \end{tikzpicture}
    \caption{Comparison of Speedup Achieved by PointSAGE in Accelerated CFD Simulations: The \textit{blue bars} represent the time taken for coarse mesh simulation along with the inference time of PointSAGE for predicting fine mesh simulation, while the \textit{red bars} represent the simulation time for fine mesh simulation using the CFD solver OpenFOAM.}
    \label{fig:simulation_time}
\end{figure}
From Figure \ref{fig:simulation_time}, it is evident that incorporating PointSAGE in the simulation process significantly reduces computation time compared to traditional CFD simulations. Specifically, simulations with PointSAGE achieve notable speedups, with a \textit{30X, 72X, and 92X} improvement in simulation time for the 2D Forward Step, 3D Lid-driven Cavity, and 2D Methane Combustion cases, respectively. These speedups are crucial for real-world applications where computational efficiency is paramount. Furthermore, PointSAGE enables accurate predictions even in unseen scenarios, as demonstrated in previous results. This dual capability positions our model as a powerful tool for accelerating computational fluid dynamics simulations while maintaining high prediction accuracy across various domains and scenarios. Due to this significant acceleration in computational speed and reliable accuracy, we can now rely on coarse simulations, thus enhancing \textit{storage and memory efficiency} in CFD simulations. Nevertheless, the model has certain \textit{limitations}. The size of the 3D point cloud we processed remains relatively modest. As this dimension expands, the "GraphSAGE" component within our model will lead to extended training periods due to the substantial volume of message-passing, thereby increasing the model's complexity. Consequently, in our future works, we aim to employ advanced versions of GNNs to tackle this challenge and guarantee the model's adaptability to larger point cloud dimensions. The framework operates under supervised learning, learning from observed coarse and fine mesh data to predict fine mesh data from unseen coarse mesh data. Future work aims to develop an unsupervised learning model for broader applicability.

\section{Conclusion}
We introduce PointSAGE, a model for superresolution using point clouds, showcasing its ability to predict fine-mesh data solely from coarse-mesh data without prior knowledge of mesh characteristics. Leveraging point cloud data, our model demonstrates robust performance across diverse datasets and unseen geometries, such as different aspect ratios and varying inlet conditions, highlighting its generalizability. The framework's adaptability enables it to predict fine mesh data of any shape/size, regardless of the dimension of the training data. In a case study of forward-facing step simulation, PointSAGE accurately captures primary shock formation and reflected shocks, achieving substantial enhancements in RMSE and MAE compared to existing deep learning techniques. Similarly, in Lid-driven cavity simulations, our model exhibits superior predictive capability in turbulent scenarios within a 3D computational domain through various scenarios, including Reynolds number extrapolation and Grid Size interpolation, with significant reductions in training time and improved MSE compared to benchmarks. The key innovation lies in our method, which eliminates the need for detailed mesh information, showcasing impressive results and setting the stage for future developments in the field. By adopting a mesh-independent approach, our work aims to revolutionize the prediction of fine-mesh simulations for fluid flows, offering a more versatile and efficient solution in CFD applications. However, the model faces limitations, particularly with the relatively small size of the 3D point cloud it handles. As the size increases, the GraphSAGE component results in longer training times due to extensive message-passing, increasing model complexity. Therefore, in future work, we plan to utilize advanced GNN versions to address this scalability issue. Also, the framework currently uses supervised learning with observed data to predict unseen fine mesh data. Future plans involve exploring unsupervised learning for broader applicability.

\section*{Impact Statement}
Our novel framework, PointSAGE, represents a significant advancement in the field of Computational Fluid Dynamics (CFD) by addressing key challenges associated with fine-mesh simulations. By leveraging the unordered, mesh-independent nature of point clouds, PointSAGE eliminates the need for intricate mesh information, making it adaptable to diverse and irregular data formats. Our model incorporates global feature extraction and local inter-dependency capture techniques, resulting in accurate predictions across a \textit{wide range of point cloud sizes, irrespective of the dimensions of the training datasets}. This capability ensures robust performance and generalization to unseen geometries, enhancing the model's utility in real-world applications.
The impact of our work is multifaceted. Firstly, PointSAGE offers a \textit{significant acceleration in computational time and reduction in memory usage} for generating fine simulations compared to traditional CFD techniques, making it an invaluable tool engineers and scientists working in fluid dynamics. Secondly, by eliminating the reliance on mesh information, our model reduces the complexity associated with data preparation and training, thereby streamlining the workflow and reducing computational resources. Lastly, the ability of PointSAGE to accurately predict fine data across diverse point cloud sizes enhances its versatility and applicability to a wide range of scenarios, from aerodynamics to environmental fluid dynamics.
Overall, PointSAGE represents a paradigm shift in CFD simulations, offering a flexible, efficient, and accurate solution to the challenges posed by fine-mesh simulations. Its impact extends beyond the realm of fluid dynamics, with potential applications in various fields requiring predictive modeling of complex systems.

\vskip 0.2in
\bibliography{reference}

\begin{thebibliography}{18}
\providecommand{\natexlab}[1]{#1}
\providecommand{\url}[1]{\texttt{#1}}
\expandafter\ifx\csname urlstyle\endcsname\relax
  \providecommand{\doi}[1]{doi: #1}\else
  \providecommand{\doi}{doi: \begingroup \urlstyle{rm}\Url}\fi

\bibitem[Bode et~al.(2019)Bode, Gauding, Kleinheinz, and Pitsch]{bode2019deep}
M.~Bode, M.~Gauding, K.~Kleinheinz, and H.~Pitsch.
\newblock Deep learning at scale for subgrid modeling in turbulent flows:
  regression and reconstruction.
\newblock In \emph{International Conference on High-Performance Computing},
  pages 541--560. Springer, 2019.

\bibitem[Bode et~al.(2021)Bode, Gauding, Lian, Denker, Davidovic, Kleinheinz,
  Jitsev, and Pitsch]{bode2021using}
M.~Bode, M.~Gauding, Z.~Lian, D.~Denker, M.~Davidovic, K.~Kleinheinz,
  J.~Jitsev, and H.~Pitsch.
\newblock Using physics-informed enhanced super-resolution generative
  adversarial networks for subfilter modeling in turbulent reactive flows.
\newblock \emph{Proceedings of the Combustion Institute}, 38\penalty0
  (2):\penalty0 2617--2625, 2021.

\bibitem[Erichson et~al.(2020)Erichson, Mathelin, Yao, Brunton, Mahoney, and
  Kutz]{erichson2020shallow}
N.~B. Erichson, L.~Mathelin, Z.~Yao, S.~L. Brunton, M.~W. Mahoney, and J.~N.
  Kutz.
\newblock Shallow neural networks for fluid flow reconstruction with limited
  sensors.
\newblock \emph{Proceedings of the Royal Society A}, 476\penalty0
  (2238):\penalty0 20200097, 2020.

\bibitem[Esmaeilzadeh et~al.(2020)Esmaeilzadeh, Azizzadenesheli, Kashinath,
  Mustafa, Tchelepi, Marcus, Prabhat, Anandkumar,
  et~al.]{esmaeilzadeh2020meshfreeflownet}
S.~Esmaeilzadeh, K.~Azizzadenesheli, K.~Kashinath, M.~Mustafa, H.~A. Tchelepi,
  P.~Marcus, M.~Prabhat, A.~Anandkumar, et~al.
\newblock Meshfreeflownet: A physics-constrained deep continuous space-time
  super-resolution framework.
\newblock In \emph{SC20: International Conference for High-Performance
  Computing, Networking, Storage and Analysis}, pages 1--15. IEEE, 2020.

\bibitem[Gao et~al.(2021)Gao, Sun, and Wang]{gao2021super}
H.~Gao, L.~Sun, and J.-X. Wang.
\newblock Super-resolution and denoising of fluid flow using physics-informed
  convolutional neural networks without high-resolution labels.
\newblock \emph{Physics of Fluids}, 33\penalty0 (7), 2021.

\bibitem[G{\"u}emes et~al.(2021)G{\"u}emes, Discetti, Ianiro, Sirmacek,
  Azizpour, and Vinuesa]{guemes2021coarse}
A.~G{\"u}emes, S.~Discetti, A.~Ianiro, B.~Sirmacek, H.~Azizpour, and
  R.~Vinuesa.
\newblock From coarse wall measurements to turbulent velocity fields through
  deep learning.
\newblock \emph{Physics of Fluids}, 33\penalty0 (7), 2021.

\bibitem[Hamilton et~al.(2018)Hamilton, Ying, and
  Leskovec]{hamilton2018inductive}
W.~L. Hamilton, R.~Ying, and J.~Leskovec.
\newblock Inductive representation learning on large graphs, 2018.

\bibitem[Hanna et~al.(2017)Hanna, Dinh, Youngblood, and
  Bolotnov]{hanna2017coarsegrid}
B.~N. Hanna, N.~T. Dinh, R.~W. Youngblood, and I.~A. Bolotnov.
\newblock Coarse-grid computational fluid dynamic (cg-cfd) error prediction
  using machine learning.
\newblock \emph{arXiv preprint arXiv:1710.09105}, 2017.

\bibitem[Hawkes et~al.(2005)Hawkes, Sankaran, Sutherland, and
  Chen]{hawkes2005direct}
E.~R. Hawkes, R.~Sankaran, J.~C. Sutherland, and J.~H. Chen.
\newblock Direct numerical simulation of turbulent combustion: fundamental
  insights towards predictive models.
\newblock In \emph{Journal of Physics: Conference Series}, volume~16, page~65.
  IOP Publishing, 2005.

\bibitem[Kim et~al.(2021)Kim, Kim, Won, and Lee]{kim2021unsupervised}
H.~Kim, J.~Kim, S.~Won, and C.~Lee.
\newblock Unsupervised deep learning for super-resolution reconstruction of
  turbulence.
\newblock \emph{Journal of Fluid Mechanics}, 910:\penalty0 A29, 2021.

\bibitem[Nair and Goza(2020)]{nair2020leveraging}
N.~J. Nair and A.~Goza.
\newblock Leveraging reduced-order models for state estimation using deep
  learning.
\newblock \emph{Journal of Fluid Mechanics}, 897:\penalty0 R1, 2020.

\bibitem[Pathak et~al.(2020)Pathak, Mustafa, Kashinath, Motheau, Kurth, and
  Day]{pathak2020using}
J.~Pathak, M.~Mustafa, K.~Kashinath, E.~Motheau, T.~Kurth, and M.~Day.
\newblock Using machine learning to augment coarse-grid computational fluid
  dynamics simulations.
\newblock \emph{arXiv preprint arXiv:2010.00072}, 2020.

\bibitem[Pfaff et~al.(2020)Pfaff, Fortunato, Sanchez-Gonzalez, and
  Battaglia]{pfaff2020learning}
T.~Pfaff, M.~Fortunato, A.~Sanchez-Gonzalez, and P.~W. Battaglia.
\newblock Learning mesh-based simulation with graph networks.
\newblock \emph{arXiv preprint arXiv:2010.03409}, 2020.

\bibitem[Qi et~al.(2017)Qi, Su, Mo, and Guibas]{qi2017pointnet}
C.~R. Qi, H.~Su, K.~Mo, and L.~J. Guibas.
\newblock Pointnet: Deep learning on point sets for 3d classification and
  segmentation.
\newblock \emph{arXiv preprint arXiv:1612.00593}, 2017.

\bibitem[Sarkar et~al.(2023)Sarkar, Majumdar, Jadhav, Sakhinana, and
  Runkana]{sarkar2023redefining}
R.~K. Sarkar, R.~Majumdar, V.~Jadhav, S.~S. Sakhinana, and V.~Runkana.
\newblock Redefining super-resolution: Fine-mesh pde predictions without
  classical simulations.
\newblock \emph{arXiv preprint arXiv:2311.09740}, 2023.

\bibitem[Xie et~al.(2018)Xie, Franz, Chu, and Thuerey]{xie2018tempogan}
Y.~Xie, E.~Franz, M.~Chu, and N.~Thuerey.
\newblock tempogan: A temporally coherent, volumetric gan for super-resolution
  fluid flow.
\newblock \emph{ACM Transactions on Graphics (TOG)}, 37\penalty0 (4):\penalty0
  1--15, 2018.

\bibitem[Yang et~al.(2019)Yang, Zhao, and Ge]{yang2019reactingfoam}
Q.~Yang, P.~Zhao, and H.~Ge.
\newblock reactingfoam-sci: An open source cfd platform for reacting flow
  simulation.
\newblock \emph{Computers \& Fluids}, 190:\penalty0 114--127, 2019.

\bibitem[Yousif et~al.(2021)Yousif, Yu, and Lim]{yousif2021high}
M.~Z. Yousif, L.~Yu, and H.-C. Lim.
\newblock High-fidelity reconstruction of turbulent flow from spatially limited
  data using enhanced super-resolution generative adversarial network.
\newblock \emph{Physics of Fluids}, 33\penalty0 (12), 2021.

\end{thebibliography}

\newpage
\appendix
\section{Appendix}

\subsection{Case Study 1: Forward Facing Step Simulation}
\label{sec:Case_study_1}
In this case study we investigate transient simulation of supersonic flow over a forward-facing step using sonicFoam solver in OpenFOAM. The sonicFoam solver is designed to solve compressible trans-sonic/supersonic laminar gas flow. The problem description involves a flow of Mach 3 at an inlet to a rectangular geometry with a step near the inlet region that generates shock waves and propagates downstream and get reflected from the walls and creates reflected shocks in the remaining length after the forward step till the time it reaches its steady state. This case study we have selected from the OpenFOAM tutorial \href{https://www.openfoam.com/documentation/tutorial-guide/3-compressible-flow/3.2-supersonic-flow-over-a-forward-facing-step}{this link}.    

\subsubsection{Problem description}
\begin{figure}[htbp]
    \centering
    \begin{tikzpicture}[scale=3]
        \draw[line width=1pt] (0,0) -- (0.6,0);
        \draw[line width=1pt] (0.6,0) -- (0.6,0.2);
        \draw[line width=1pt] (0.6,0.2) -- (3,0.2);
        \draw[line width=1pt] (3,0.2) -- (3,1);
        \draw[line width=1pt] (3,1) -- (0,1);
        \draw[line width=1pt] (0,1) -- (0,0);
        
        \draw[-{Latex[length=5mm,width=1.5mm]}, line width=2pt] (-0.8,0.5) -- (-0.2,0.5) node[midway,above] {$U_{\infty}$};
        
        \draw[|<->|] (0,-0.2) -- (0.6,-0.2) node[midway,below] {0.6m};
        \draw[|<->|] (3.1,0.2) -- (3.1,1) node[midway,below, rotate=90] {0.8m};
        \draw[|<->|] (0.7,0) -- (0.7,0.2) node[midway,right] {0.2m};
        \draw[|<->|] (0.6,-0.2) -- (3,-0.2) node[midway,below] {2.4m};
        \draw[|<->|] (-0.1,0) -- (-0.1,1) node[midway,above,rotate=90] {1m};
        
        \node[rotate=90] at (2.9,0.6) {Wall};
        \node at (1.5,1.1) {Wall};
    \end{tikzpicture}
    \caption{Computational Domain of a 2D Forward Facing Step Simulation}
    \label{fig:Domain}
\end{figure}
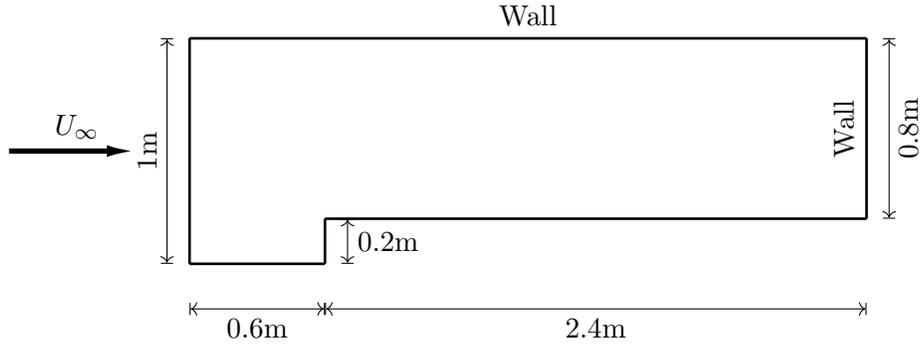

\textbf{Solution domain}\\
The 2D computational domain features a step with a height of 20\% located at a distance of 0.6m from the inlet as shown in the Figure \ref{fig:Domain}. The experiment is conducted in a gas medium with a speed of sound given by \(\sqrt{\gamma R T} = 1\) m/s. Thus, at the inlet, the flow is supersonic with a Mach number of 3 (\(U_\infty = 3\) m/s), along with a pressure of 1 Pa and a temperature of 1 K.
The aspect ratio of the defined geometry in this case study is expressed as the ratio of the length of the rectangular domain (3m) to its height (1m), as shown below:

\[ \text{Aspect ratio} = \frac{\text{Length}}{\text{Height}} = \frac{3 \, \text{m}}{1 \, \text{m}} \]

\textbf{Governing equations}

Mass continuity:
\begin{equation}
    \frac{\partial \rho}{\partial t} + \nabla \cdot (\rho \mathbf{U}) = 0
\end{equation}

Ideal gas:
\begin{equation}
    p = \rho R T
\end{equation}

Momentum equation for Newtonian fluid:
\begin{equation}
    \frac{\partial (\rho \mathbf{U})}{\partial t} + \nabla \cdot (\rho \mathbf{U} \mathbf{U}) - \nabla \cdot \mu \nabla \mathbf{U} = - \nabla p
\end{equation}

The energy equation for fluid (ignoring some viscous terms):
\begin{equation}
    \frac{\partial (\rho e)}{\partial t} + \nabla \cdot (\rho \mathbf{U} e) - \nabla \cdot \left( \frac{k}{C_v} \nabla e \right) = p \nabla \cdot \mathbf{U}
\end{equation}

\textbf{Initial Conditions:}
\begin{align*}
    & U = 0 \, \mathrm{m/s}, \quad p = 1 \, \mathrm{Pa}, \quad T = 1 \, \mathrm{K}
\end{align*}

\textbf{Boundary Conditions:}
\begin{itemize}
    \item \textbf{Inlet (left):}
    \begin{align*}
        & \text{FixedValue for velocity: } U = 3 \, \mathrm{m/s} \, (\text{Mach 3}) \\
        & \text{Pressure: } p = 1 \, \mathrm{Pa} \\
        & \text{Temperature: } T = 1 \, \mathrm{K}
    \end{align*}
    
    \item \textbf{Outlet (right):}
    \begin{align*}
        & \text{ZeroGradient on } U, p, \text{ and } T
    \end{align*}
    
    \item \textbf{No-slip adiabatic wall (bottom)}
    
    \item \textbf{Symmetry plane (top)}
\end{itemize}

\textbf{Transport Properties:}
\begin{align*}
    & Laminar 
    & \text{Dynamic viscosity of air: } \mu = 18.1 \, \mu\mathrm{Pas}
\end{align*}

\textbf{Thermodynamic Properties:}
\begin{align*}
    & \text{Specific heat at constant volume: } C_v = 1.78571 \, \mathrm{J/kgK} \\
    & \text{Gas constant: } R = 0.714286 \, \mathrm{J/kgK} \\
    & \text{Conductivity: } k = 32.3 \, \mu\mathrm{W/mK}
\end{align*}

\subsubsection{Mesh Description}
The mesh is generated using the \textit{blockMesh} utility, dividing the domain into uniform rectangular cells. For the fine mesh, the cells have dimensions of 0.03 m in the \(x\)-direction and 0.025 m in the \(y\)-direction, resulting in 42702 points for the point cloud. Conversely, the coarse mesh divides the domain into cells with dimensions of 0.12 m in the \(x\)-direction and 0.1 m in the \(y\)-direction, yielding 3352 points for the coarse point cloud used in our neural network.

\begin{figure}[htbp]
    \centering
    \begin{subfigure}[b]{0.45\textwidth}
        \centering
        \includegraphics[width=1.1\textwidth, trim={14cm 12cm 14cm 12cm}, clip]{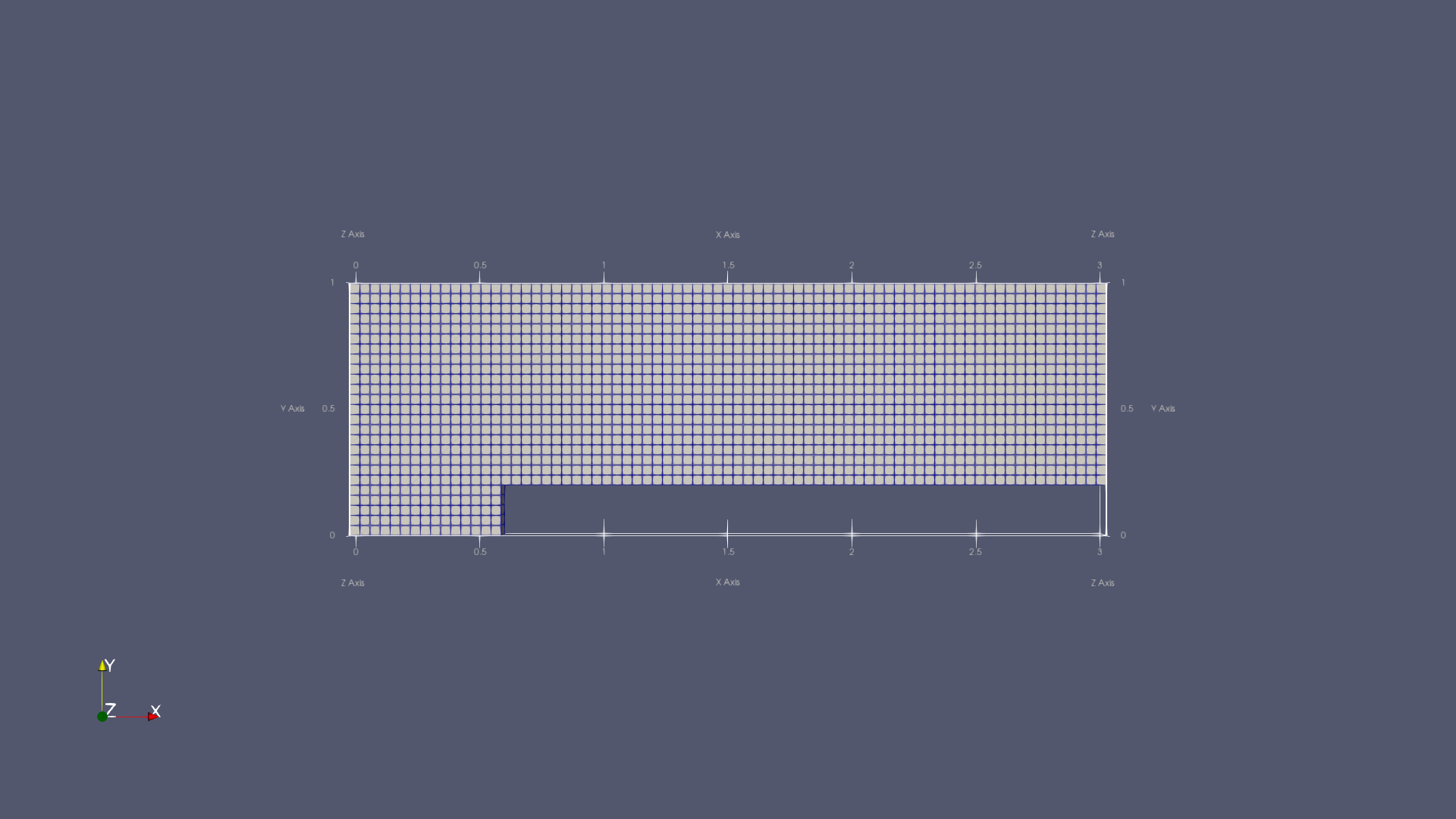}
        \caption{Coarse Mesh}
        \label{fig:coarse_mesh}
    \end{subfigure}
    \hspace{0.75cm}
    \begin{subfigure}[b]{0.45\textwidth}
        \centering
        \includegraphics[width=1.05\textwidth, trim={8cm 8cm 5cm 10cm}, clip]{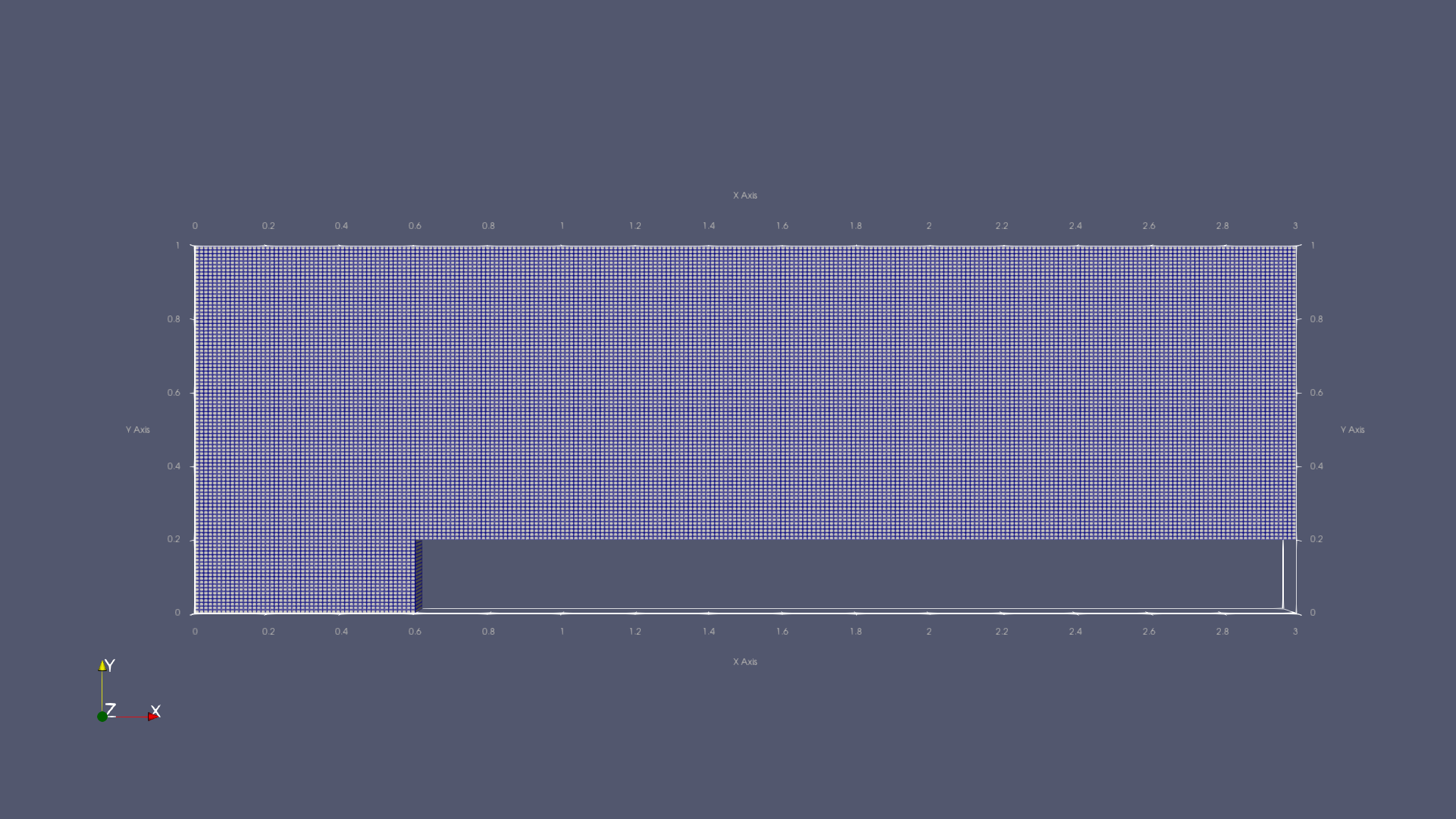}
        \caption{Fine Mesh}
        \label{fig:fine_mesh}
    \end{subfigure}
    \caption{Mesh Description: Aspect Ratio 3}
    \label{fig:mesh_description}
\end{figure}

\subsubsection{Scenario 1}
\begin{figure}[htbp]
    \centering
    \includegraphics[width =\textwidth]{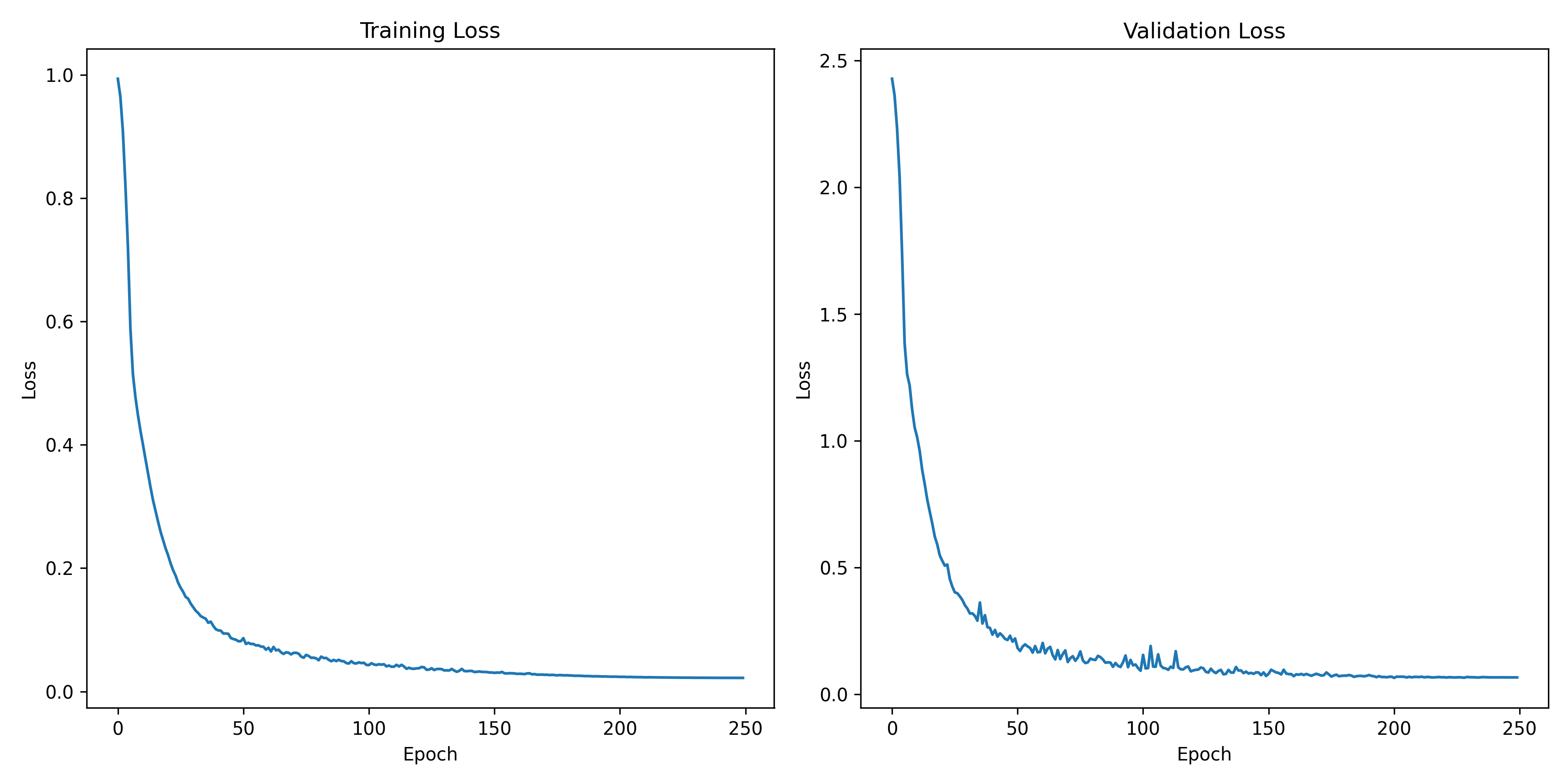}
    \caption{PointSAGE Training and Validation Loss}
    \label{fig:loss}
\end{figure}

\begin{figure}[htbp]
    \centering
    \begin{subfigure}[b]{\textwidth}
        \centering
        \includegraphics[width=1.03\linewidth, trim={0.2cm 0.2cm 0cm 0cm}, clip]{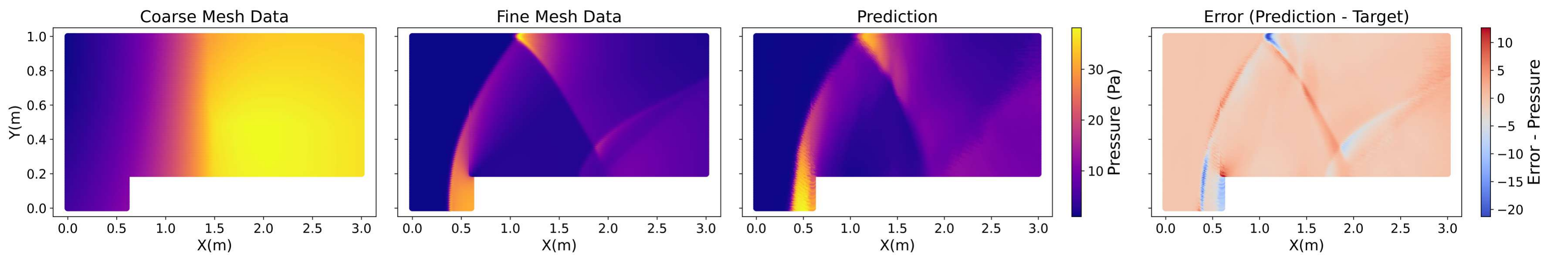}
        \caption{time-step \(t = 2s\)}
        \label{fig:sub1}
    \end{subfigure}

    \begin{subfigure}[b]{\textwidth}
        \centering
        \includegraphics[width=1.03\linewidth, trim={0.2cm 0.2cm 0cm 0cm}, clip]{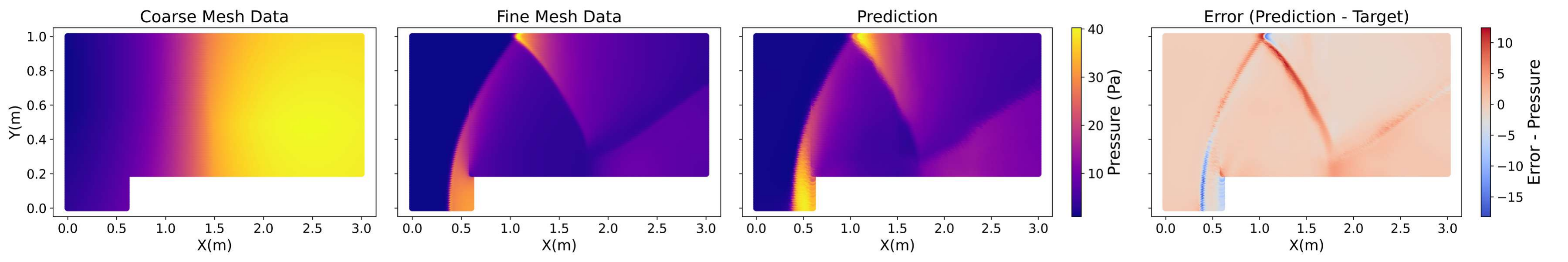}
        \caption{time-step \(t = 4.5s\)}
        \label{fig:sub2}
    \end{subfigure}

    \begin{subfigure}[b]{\textwidth}
        \centering
        \includegraphics[width=1.03\linewidth, trim={0.2cm 0.2cm 0cm 0cm}, clip]{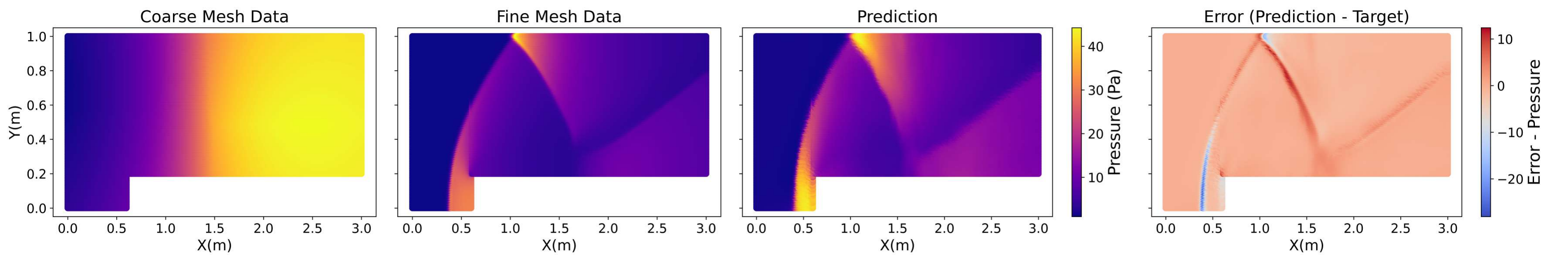}
        \caption{time-step \(t = 7s\)}
        \label{fig:sub3}
    \end{subfigure}

    \begin{subfigure}[b]{\textwidth}
        \centering
        \includegraphics[width=1.03\linewidth, trim={0.2cm 0.2cm 0cm 0cm}, clip]{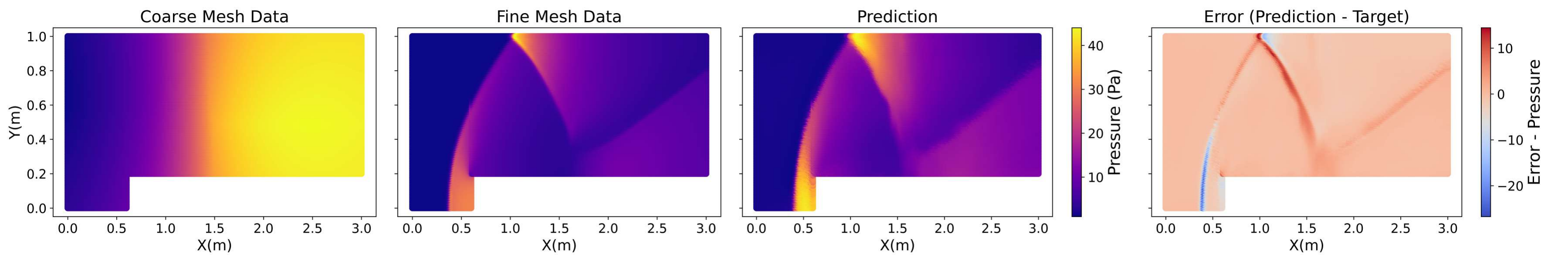}
        \caption{time-step \(t = 9.5s\)}
        \label{fig:sub4}
    \end{subfigure}

    \caption{Pressure distribution from PointSAGE-predicted fine mesh simulation at various time steps, corresponding to an inlet velocity \(U_\infty\) of 4.875 m/s.}
    \label{fig:pressure_scenario_1}
\end{figure}

\begin{figure}[htbp]
    \centering
    \begin{subfigure}[b]{\textwidth}
        \centering
        \includegraphics[width=1.03\linewidth, trim={0.2cm 0.2cm 0cm 0cm}, clip]{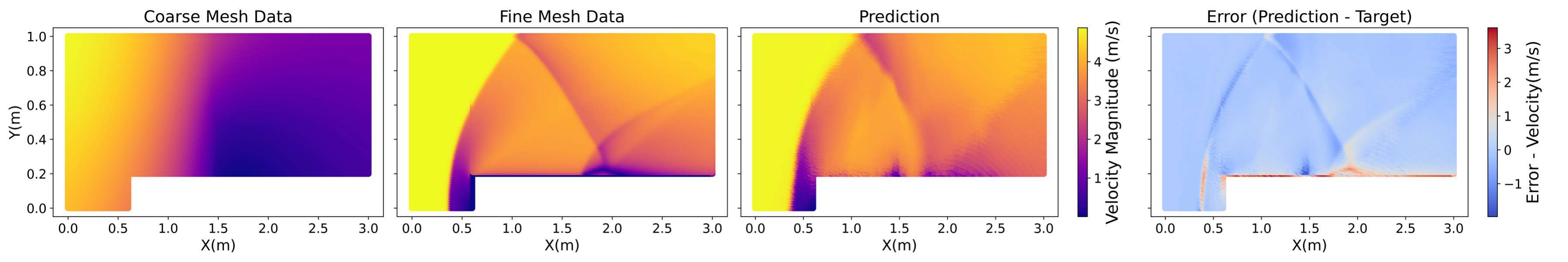}
        \caption{time-step \(t = 2s\)}
        \label{fig:sub5}
    \end{subfigure}

    \begin{subfigure}[b]{\textwidth}
        \centering
        \includegraphics[width=1.03\linewidth, trim={0.2cm 0.2cm 0cm 0cm}, clip]{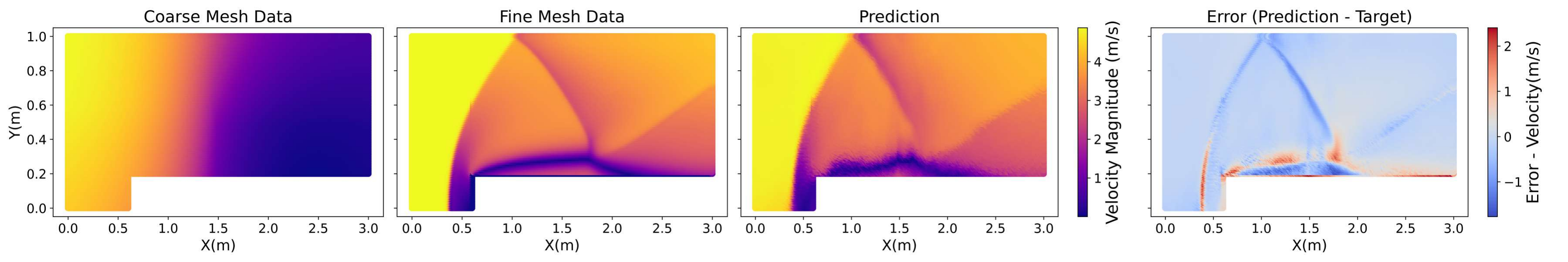}
        \caption{time-step \(t = 4.5s\)}
        \label{fig:sub6}
    \end{subfigure}

    \begin{subfigure}[b]{\textwidth}
        \centering
        \includegraphics[width=1.03\linewidth, trim={0.2cm 0.2cm 0cm 0cm}, clip]{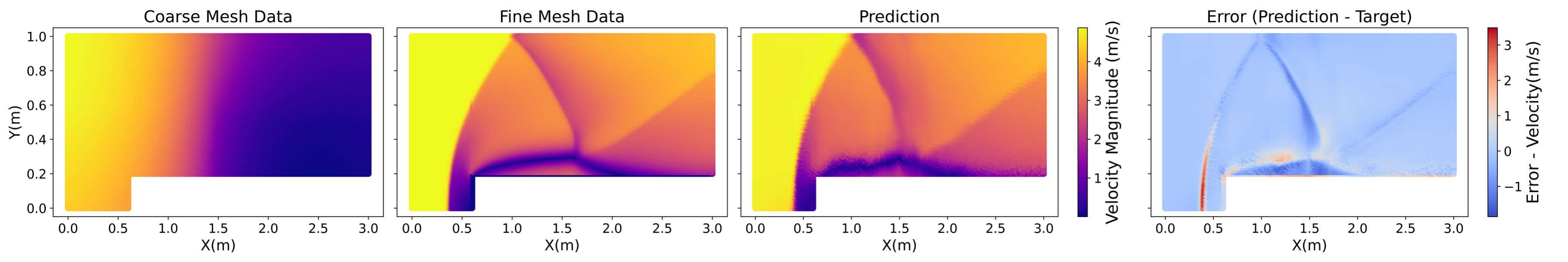}
        \caption{time-step \(t = 7s\)}
        \label{fig:sub7}
    \end{subfigure}

    \begin{subfigure}[b]{\textwidth}
        \centering
        \includegraphics[width=1.03\linewidth, trim={0.2cm 0.2cm 0cm 0cm}, clip]{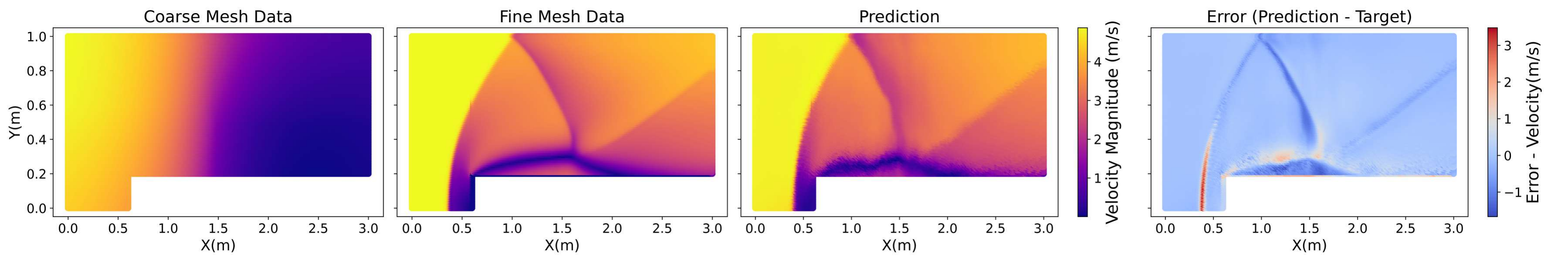}
        \caption{time-step \(t = 9.5s\)}
        \label{fig:sub8}
    \end{subfigure}

    \caption{Velocity distribution from PointSAGE-predicted fine mesh simulation at various time steps, corresponding to an inlet velocity \(U_\infty\) of 4.875 m/s.}
    \label{fig:velocity_scenereo_1}
\end{figure}

In this scenario, we conducted experiments using a dataset where the inlet velocity was varied in the range of 2 m/s to 5 m/s, with intervals of 0.25 m/s, while maintaining a constant aspect ratio of 3, as mentioned earlier. The simulations were performed on an Intel(R) Core(TM) i7-8700 CPU @ 3.20GHz. For deep learning experiments, we partitioned the dataset into 80\%/10\%/10\% for training, validation, and testing respectively, and executed the entire experiment on a Tesla P100 GP with 16GB VRAM.\\
\textbf{Results}\\
The simulations conducted were of a transient nature, and PointSAGE demonstrated commendable accuracy in predicting features such as pressure and velocity at different time intervals, as illustrated in Figure \ref{fig:pressure_scenario_1} and Figure \ref{fig:velocity_scenereo_1} respectively. Both figures reveal that our PointSAGE model effectively captures the propagation of shocks and their reflection within the rectangular domain following the step location. The training and validation for the PointSAGE training can be observe in this Figure \ref{fig:loss}. 

\subsubsection{Scenario 2:}
\begin{figure}[htbp]
    \centering
    \includegraphics[width=1.03\linewidth, trim={0.2cm 0.2cm 0cm 0cm}, clip]{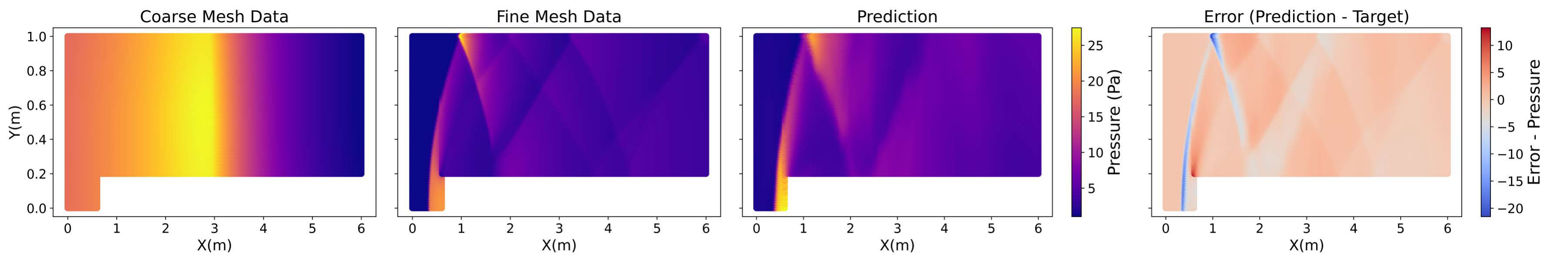}
    \includegraphics[width=1.03\linewidth, trim={0.2cm 0.2cm 0cm 0cm}, clip]{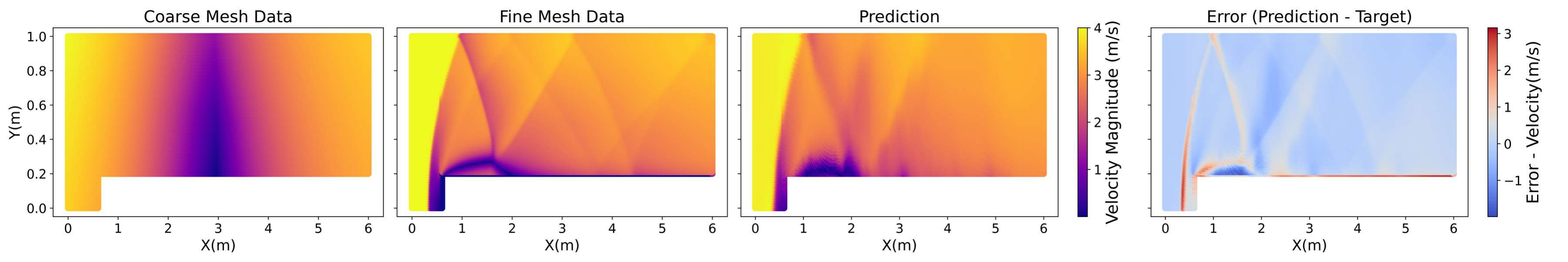}
    \caption{Scenario 2: Training on AR 3 and 4,and pressure and velocity prediction at t= 3.5s for inlet velocity 4 m/s for an AR 6 dataset.}
    \label{fig:Scenario_2}
\end{figure}

In this particular scenario, we conducted experiments utilizing a dataset wherein the inlet velocity was systematically varied within the range of 2 m/s to 5 m/s, with intervals of 0.5 m/s. Additionally, we varied the aspect ratio within the range of 3 to 6. The variation in aspect ratio involves an increase in the length after the step location at 0.6 m. For instance, in the case of an aspect ratio of 3, the length of the section after the step is 2.4 m (total length = 0.6 + 2.4 = 3m). On the other hand, for an aspect ratio of 4, the length of the section after the step is 3.4 m (total length = 0.6 + 3.4 = 4m). In the context of deep learning experiments, we partitioned the dataset for training using aspect ratios 3 and 4, for validation with aspect ratio 5, and for testing with aspect ratio 6.\\
\textbf{Results}\\
The objective of this scenario is to evaluate the model's proficiency in effectively understanding and adapting to physical phenomena, specifically shock formation and reflection, within a given aspect ratio. Furthermore, the model is challenged to extend its predictions to another aspect ratio, adding complexity as an increase in the length after the step leads to intensified shock reflection and sudden alterations in flow behavior downstream.
As illustrated in Figure \ref{fig:Scenario_2}, PointSAGE demonstrates satisfactory predictions for essential features such as pressure and velocity. This success underscores the model's ability to effectively capture and forecast the dynamic behaviors of shocks under varying aspect ratios, emphasizing its efficacy in handling complex flow phenomena.

\subsection{Case Study 2: Lid-driven Cavity}
In this case study, we explore the lid-driven cavity, a relatively straightforward scenario compared to others, well known as a benchmark problem in computational fluid dynamics (CFD). The problem entails modeling the fluid flow within a cubic cavity, with a width of 1m and a lid velocity (\(U_{lid}\)) of 1m/s, resulting in intricate fluid phenomena, notably the formation of counter-rotating vortices at the cavity's bottom. The flow characteristics vary depending on factors such as the Reynolds number and aspect ratio. To simulate this, we employed a transient solver known as \textit{pisoFoam}, implementing the PISO algorithm. For the present work, it is focused on the quasi-steady state flow within the cavity.

\begin{figure}[htbp]
    \centering
    \includegraphics[width=0.25\linewidth]{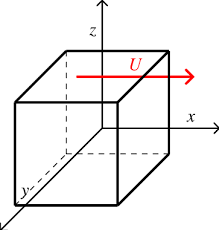}
    \caption{Lid Driven Cavity}
    \label{fig:Lid-Drive-Cavity}
\end{figure}

\subsubsection{Problem Description}
\begin{figure}[htbp]
\centering
\subfloat[Coarse Mesh]{
\includegraphics[width=0.2\linewidth]{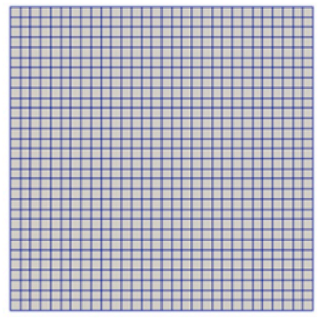} 
}
\subfloat[Fine Mesh along with wall refinement]{
 \includegraphics[width=0.65\linewidth]{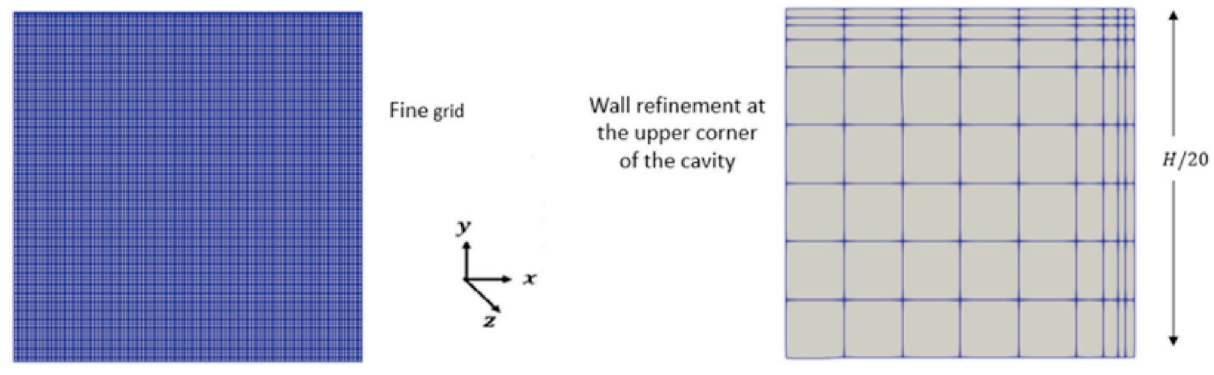}
}
\caption{The above figures depict the computational grids utilized in this work}
\label{Mesh information}
\end{figure}

As explained above, the computational domain comprises a 3D cube cavity with a width of 1m, as illustrated in Figure \ref{fig:Lid-Drive-Cavity}. The cavity's height adjusts proportionally based on the aspect ratio. In this study, varying Reynolds numbers are achieved by altering the kinematic viscosity while maintaining a constant \(U_{lid}\) value. Specifically, Reynolds numbers of 6000, 8000, 10000, and 12000 are considered, with grid sizes of 1/20, 1/30, and 1/40 employed for simulating coarse mesh data. For fine mesh data, a grid size of 1/120 is utilized. To enhance turbulence capture within the cavity, wall refinement is implemented in the fine mesh data's wall region, as depicted in the Figure \ref{Mesh information}, whereas such refinement is omitted for coarse data.

Following the generation of fine mesh data, it is overlaid onto the coarse mesh, resulting in identical mesh sizes denoted as \((r, d)\), where \(r\) signifies the number of points, and \(d\) represents the number of features. In this specific instance, the model's up-sampling aspect is bypassed since the input and output sizes match. In alignment with the source paper's methodology, which evaluated the model's predictability across diverse scenarios, including Reynolds number extrapolation, our study adheres to a similar approach. The model is trained on a subset of Reynolds numbers (e.g., 6000, 80000, 10000) and tested on entirely different ones (e.g., 12000), replicating six such scenarios. Herein, we concentrate on the initial four scenarios to showcase our model's versatility.

\textbf{Results:} To demonstrate the model's effectiveness and predictive ability, we adhere to the protocol outlined in the source paper (\cite{hanna2017coarsegrid}). In the six scenarios, we will execute the first four scenarios. The \textbf{"Scenario-1" (Reynolds number interpolation)} was showcased in the main paper. In this section, we will delve into the other scenarios. In these scenarios, the model's hyper-parameters consist of \textbf{r}, representing the radius of the graph, and \textbf{lr}, indicating the learning rate. Following several iterations, it was observed that the most effective values for "k" and "lr" are 0.005 and 0.001, respectively. Furthermore, each model undergoes training for 300 epochs.

\begin{figure}[htbp]
    \centering
    \includegraphics[width=\linewidth, trim={0.2cm 2cm 0cm 2cm}, clip]{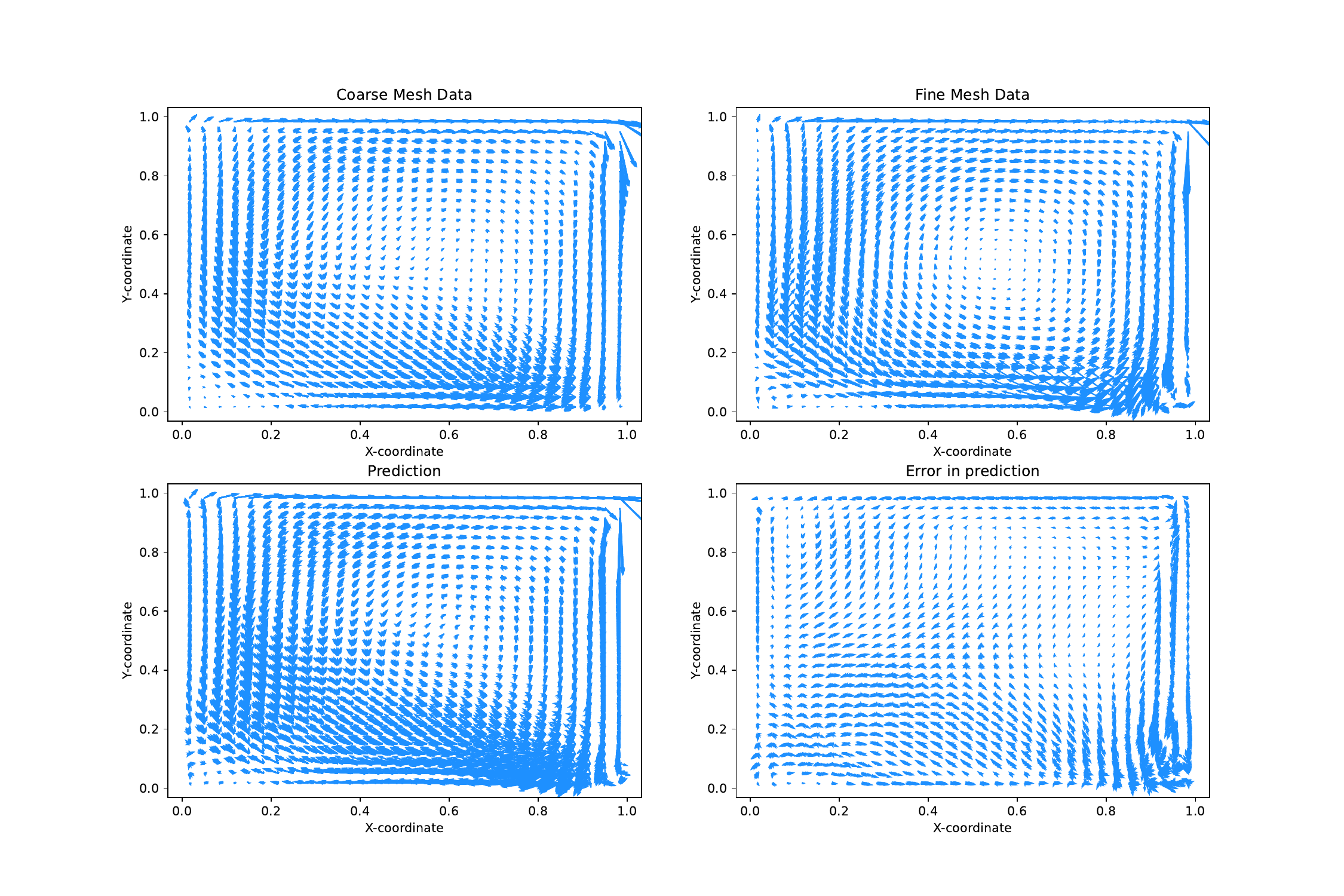}
    \caption{Velocity Contour Plot(\(U_x\) and \(U_y\)): Scenario - 2}
    \label{fig:Lid-Drive-Cavity2}
\end{figure}

\begin{figure}[htbp]
    \centering
    \includegraphics[width=\linewidth, trim={0.2cm 15cm 0cm 3cm}, clip]{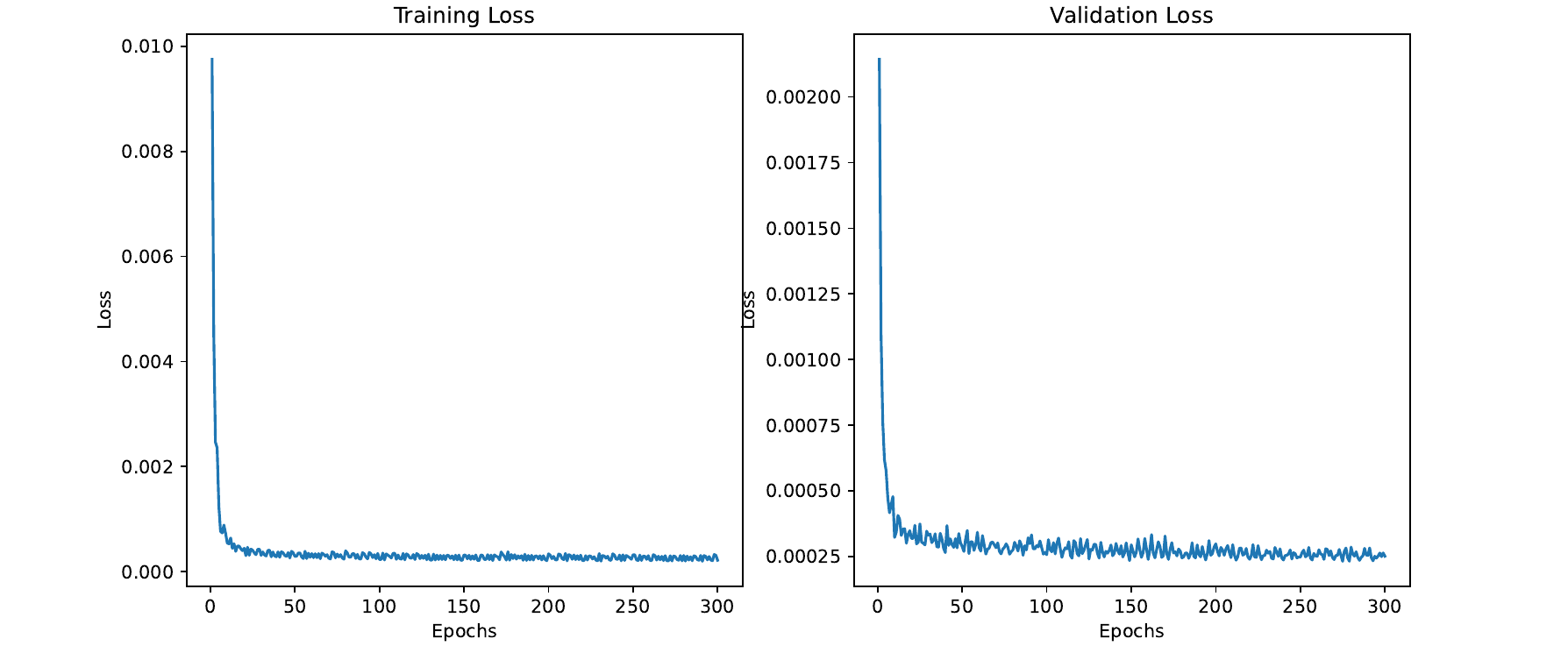}
    \caption{Training and Validation Loss}
    \label{fig:Loss2}
\end{figure}

\textbf{Scenario - 2:} This scenario illustrates Reynolds number extrapolation, where training and validation are conducted for flow at \textbf{Re - {6000, 8000, 10000}}, and the model is subsequently tested at a different Reynolds number, \textbf{12000}. The grid size (1/30) and aspect ratio (1) remain constant. Figure \ref{fig:Lid-Drive-Cavity2} showcases the model's predictive prowess, with velocity contours plotted. The model accurately captures the turbulent nature of the flow, achieving an MSE of 3.5e-4 within just 120 seconds. Figure \ref{fig:Loss2} showcase the training and validation loss.

In this case, since the grid size and the aspect ratio remain constant, the number of points in the input/output point cloud remains constant, i.e., 27,000. However, if we alter either of them, the number of points changes. In the next two scenarios, the grid size is modified, leading to different point cloud dimensions for the training and testing datasets.

\begin{figure}[htbp]
    \centering
    \includegraphics[width=\linewidth]{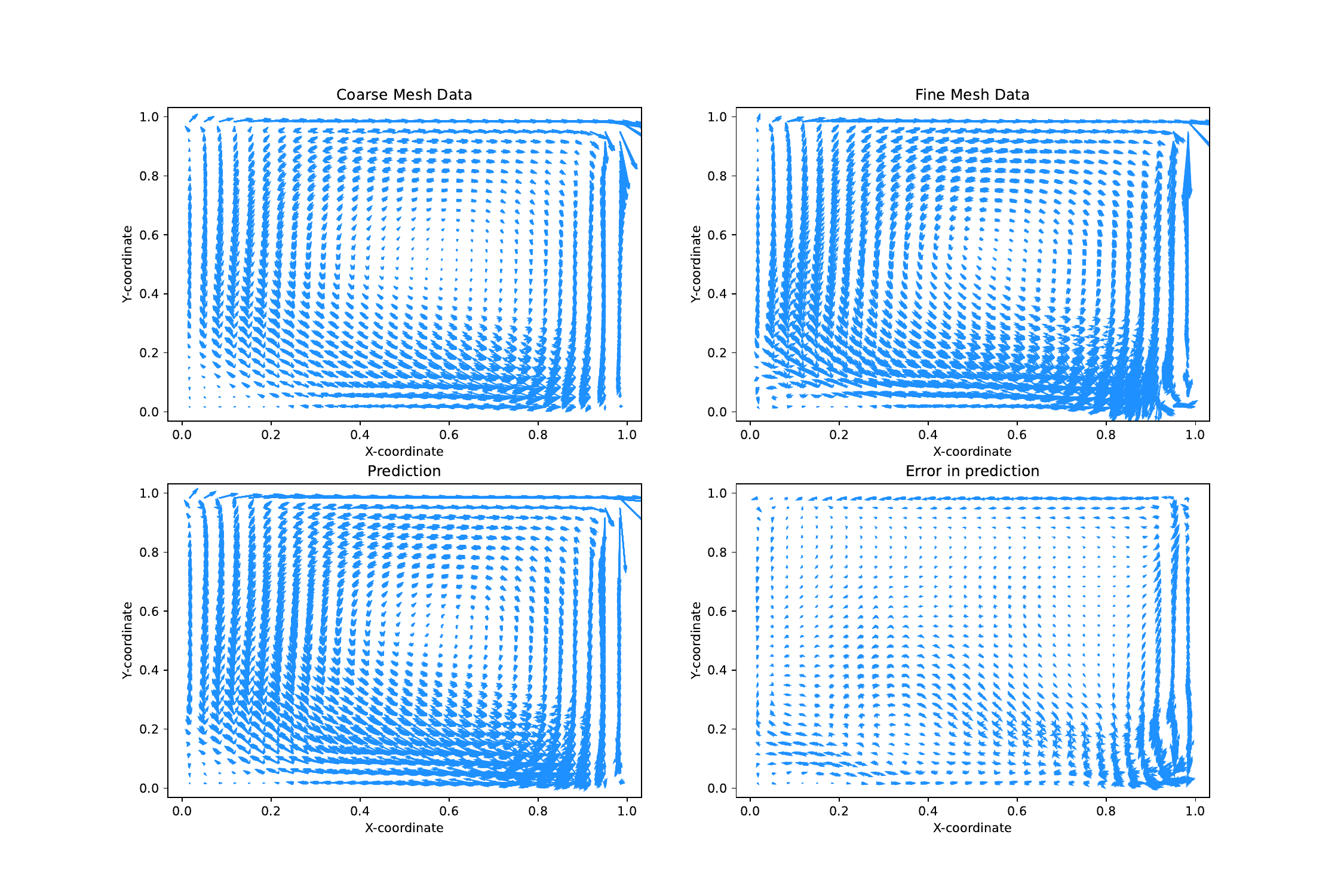}
    \caption{Velocity Contour Plot(\(U_x\) and \(U_y\)): Scenario - 3}
    \label{fig:Lid-Drive-Cavity3}
\end{figure}

\begin{figure}[htbp]
    \centering
    \includegraphics[width=\linewidth, trim={0.2cm 15cm 0cm 3cm}, clip]{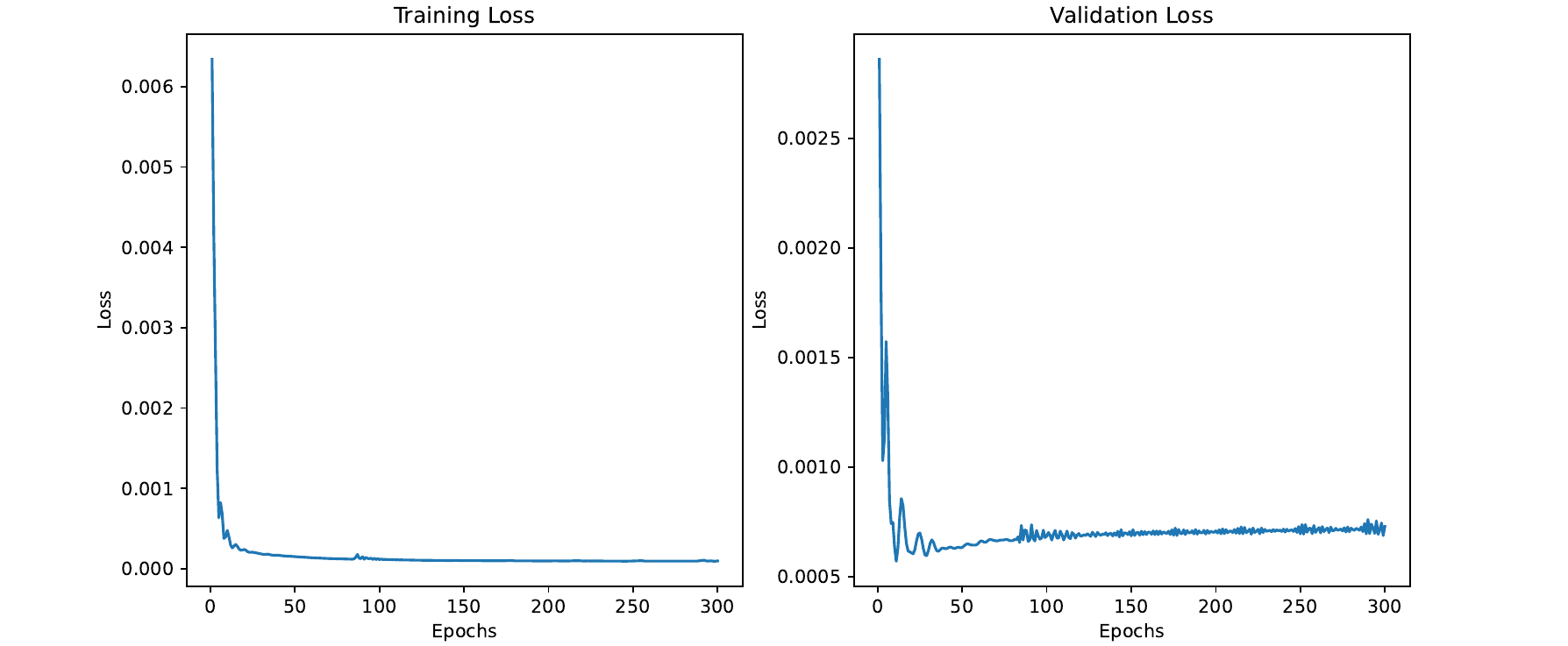}
    \caption{Training and Validation Loss}
    \label{fig:Loss3}
\end{figure}

In \textbf{Scenario - 3}, Reynolds number and Grid Size interpolation are employed, where training and validation encompass flow conditions at \textbf{Re - {8000, 12000} and Grid size - {1/40, 1/20}}. The model is then assessed with a different parameter set, \textbf{Re - 10000 and Grid Size - 1/30}. The point cloud dimensions for training, validation, and testing datasets are 64,000, 8,000, and 27,000, respectively, showcasing the model's adaptability to various mesh dimensions. Figure \ref{fig:Lid-Drive-Cavity3} illustrates the velocity contour plot, highlighting the model's adeptness in accurately capturing turbulence, achieving an MSE of 3e-4 within a mere 156 seconds. Figure \ref{fig:Loss3} showcase the training and validation loss.

\begin{figure}[htbp]
    \centering
    \includegraphics[width=\linewidth, trim={0.2cm 15cm 0cm 3cm}, clip]{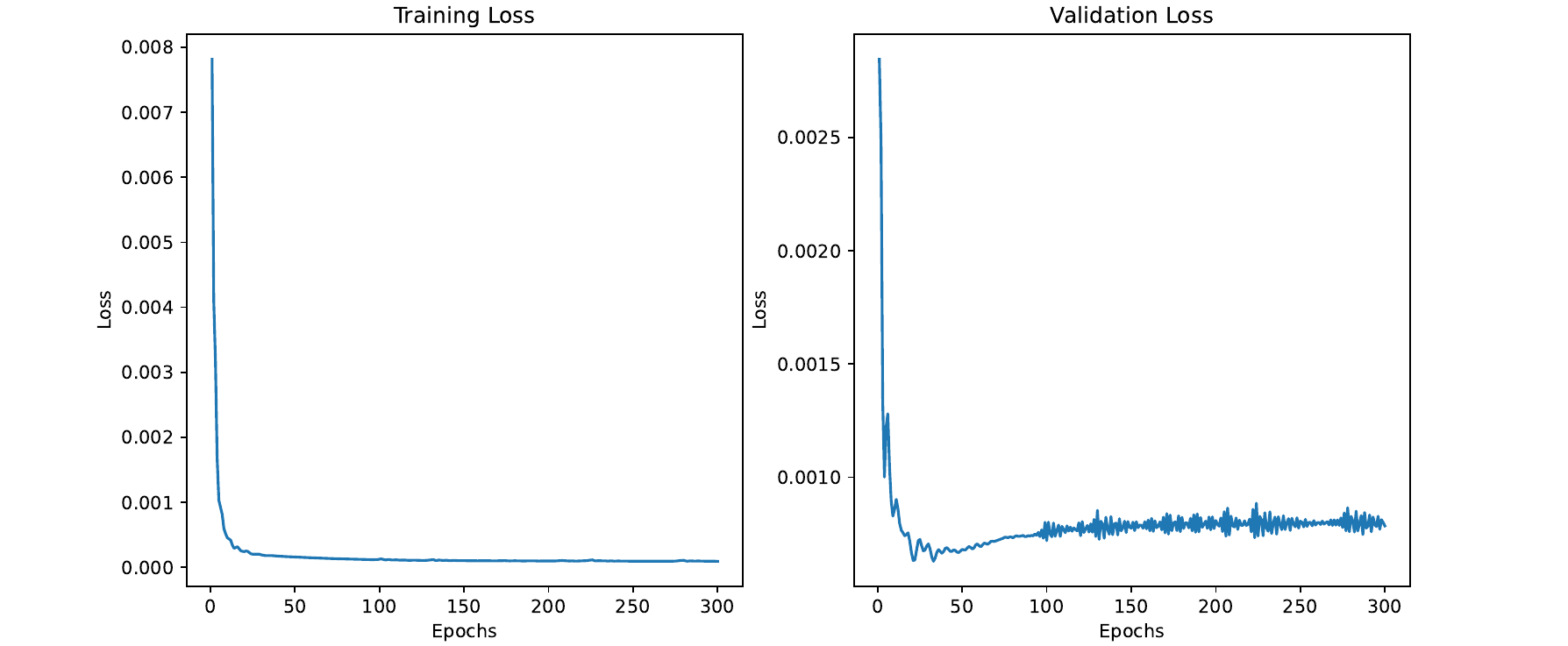}
    \caption{Training and Validation Loss}
    \label{fig:Loss4}
\end{figure}

\begin{figure}[htbp]
    \centering
    \includegraphics[width=\linewidth]{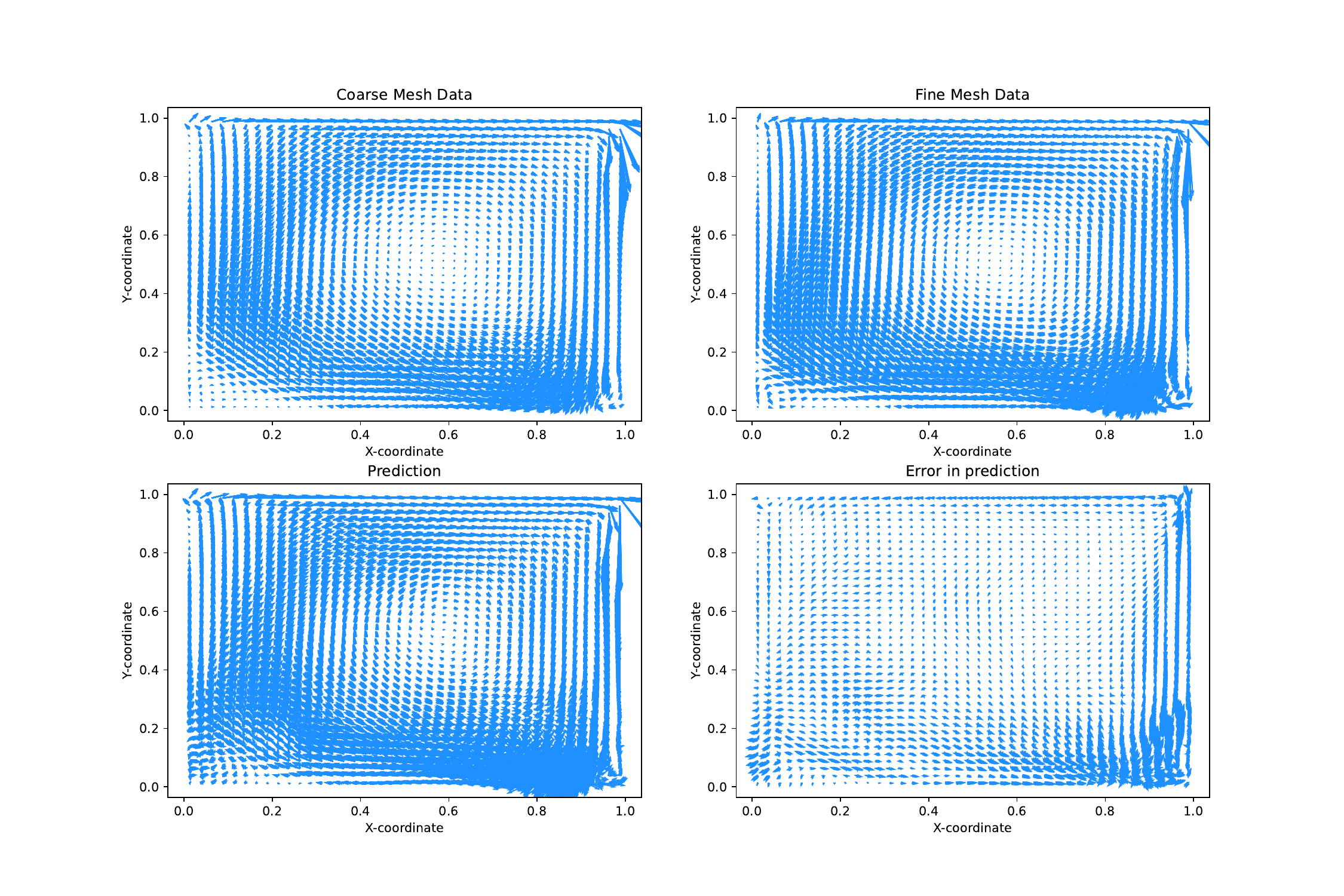}
    \caption{Velocity Contour Plot(\(U_x\) and \(U_y\)): Scenario - 4}
    \label{fig:Lid-Drive-Cavity4}
\end{figure}

\textbf{Scenario - 4:} In this scenario, both Reynolds number and grid size are varied for training and validation, covering flow conditions at \textbf{Re - {8000, 10000} and Grid size - {1/30, 1/20}}. The model is then tested with a different parameter set, \textbf{Re - 12000 and Grid Size - 1/40}. The point cloud dimensions for training, validation, and testing datasets are 64,000, 8,000, and 27,000, respectively. This scenario presents a greater challenge compared to the previous one. Figure \ref{fig:Lid-Drive-Cavity4} demonstrates the model's efficacy in accurately capturing turbulence with an MSE of 2e-4 within just 50 seconds. Figure \ref{fig:Loss4} showcase the training and validation loss.

\end{document}